\documentclass[12pt,a4paper]{article}
\usepackage{amsmath}
\usepackage{slashed}
\usepackage{amssymb}
\usepackage[usenames, dvipsnames]{color}
\usepackage{mathtools}
\usepackage{verbatim}
\usepackage{amsfonts}
\usepackage{mathrsfs}
\usepackage{graphicx}
\usepackage[font=small]{caption}
\usepackage[caption=false]{subfig}
\usepackage{enumitem}
\usepackage[normalem]{ulem}
\usepackage[percent]{overpic}
\usepackage{bm}
\usepackage{bbm}
\usepackage{rotating}
\usepackage{hyperref}
\usepackage{cancel}
\usepackage{verbatim}
\usepackage{graphicx}
\usepackage{tensor}
\usepackage{wrapfig}
\usepackage{booktabs}
\usepackage{natbib}
\usepackage{setspace}
\usepackage{relsize}
\usepackage{authblk}
\usepackage{multirow}

\usepackage[hang,flushmargin]{footmisc}
\AtBeginEnvironment{footnote}{\footnotesize}

\usepackage{geometry}
\usepackage{braket}
\newcommand{\ii}{\mathrm{i}}

\renewcommand{\d}{\mathrm{d}}

\newcommand{\GG}[1]{}

\begin{document}
\title{\Large Spacetime Representation Theory: Setting the Scope\\ of the ISE Method of Topological Redescription}
\author{\vspace{-0.25cm} \smaller Daniel Grimmer}\footnotetext{\ \ Email: daniel.grimmer@philosophy.ox.ac.uk\\
\ \ Orchid ID: 0000-0002-8449-3775}
\affil{\small \vspace{-0.25cm} Faculty of Philosophy, University of Oxford, Oxford, United Kingdom\\
Reuben College, University of Oxford, Oxford, United Kingdom}
\date{}

{\singlespacing \maketitle}
\begin{abstract}
Spacetime dualities arise whenever two theories---despite being structurally equivalent in some sense---seemingly provide us with two radically different spatiotemporal descriptions of the world. This often involves radical differences in how the two theories topologically stage their states; Whereas one theory is about \textit{this} type of particle/field on \textit{this} smooth manifold, the other theory is about \textit{that} type of particle/field arranged differently on \textit{that} smooth manifold. For instance, the AdS-CFT correspondence relates a certain theory set in the bulk (our 3+1 dimensional spacetime) to another theory set on the boundary (a 2+1 dimensional spacetime). Another example (new in this paper) is the M\"{o}bius-Euclid duality: a theory about a certain type of particle floating around on the Euclidean plane can be topologically redescribed as instead being about a different type of particle living on a M\"{o}bius strip, and vice versa. 

The possibility of such alternative spacetime framings raises some significant questions about the epistemology and metaphysics of space and time. For instance, what are our topology selection criteria? Are they objective or conventional? Moreover, given that two spacetime theories are topological redescriptions of each other, what is the common core which they are equivalent descriptions of? As a step towards answering such questions, this paper develops a general framework (spacetime representation theory) for understanding our ability to topologically redescribe our spacetime theories. With this framework established, I will then discuss the ISE Equivalence Theorem which sets the scope of the recently developed ISE Method of topological redescription.
\end{abstract}

\newgeometry{lmargin=1.5in, rmargin=1.5in, tmargin=1.0in, bmargin=1.0in}

\section{Introduction and Motivation}\label{SecIntroduction}
\quad \ \, Spacetime dualities arise when two or more theories---despite being structurally equivalent in some sense---seemingly provide us with two radically different spatiotemporal descriptions of the world. This often involves radical differences in how the two theories topologically stage their states; Whereas one theory is about \textit{this} type of particle/field on \textit{this} smooth manifold, the other theory is about \textit{that} type of particle/field arranged differently on \textit{that} smooth manifold. An illustrative example drawn from the front lines of physics is the AdS-CFT correspondence in which a theory about a certain type of field living in the bulk (our 3+1 dimensional spacetime) has an equivalent description in terms of some other type of field living on the boundary (a 2+1 dimensional spacetime). In Sec.~\ref{SecFirstTopRed}, I will introduce the M\"{o}bius-Euclid duality: a theory about a certain type of particle floating around on a Euclidean plane can be topologically redescribed as instead being about a different type of particle living on the Möbius strip, and vice versa. 

The possibility of such alternative spacetime framings raises some significant questions about the epistemology and metaphysics of space and time.\footnote{There have been a great many papers have been written about the philosophical consequences of spacetime dualities. See, for instance, the work of De Haro and Butterfield \cite{de_haro_comparing_2017,de_haro_schema_2018,de_haro_theoretical_2019,de_haro_symmetry_2021,de_haro_empirical_2023,deharo2023} among many others \cite{matsubara_realism_2013,huggett_emergent_2013,huggett_target_2017,rickles_dual_2017,le_bihan_duality_2018,read_motivating_2018,jaksland_holography_2020,RasmusEnrico}.\label{FnDualityWorks}} It raises questions about our selection criteria for spacetime's topological properties (e.g., its connectedness, continuity, dimensionality, etc.). Past this, further selection criteria may then be needed to decide how our theories' states are to be staged upon a given topological backdrop (see Sec.~\ref{SecMoreTopRed}). But what are these topology selection criteria exactly? Are these criteria objective? Or is there some element of conventionality to them? For instance, in a M\"{o}bius-Euclid world, could it be that one rational civilization opts for the M\"{o}bius representation whereas another opts for the Euclid representation? Moreover, given that two spacetime theories are topological redescriptions of each other, what is the common core which they are equivalent descriptions of? Does this common structure provide a sufficiently rich ontology?

As I have argued in my recent DPhil thesis \citep{GrimmerThesis}, our efforts to answer these questions would be greatly enhanced if we could gain some mathematical control over the range of all possible spacetime representations of our theories.\footnote{This manuscript is an adaptation of various chapters of my thesis. Specifically, it pulls material from \citeauthor{GrimmerThesis} (\citeyear[Chs. 3 - 6]{GrimmerThesis}).} Ideally, we would be able to remove and replace the topological underpinnings of our spacetime theories just as easily as we can switch between different coordinate systems (or different law-like axiomatizations, or different geometric framings). \cite{GrimmerThesis} has shown that we can, in fact, do so  by using the newly developed ISE Method of topological redescription. The ultimate scope of the ISE Method is set by the ISE Equivalence Theorem. In effect, this theorem states that the ISE Method can be applied to a very wide scope of spacetime theories and gives us access to effectively every possible topological redescription thereof; Its range is only limited by a weak spacetime-kinematic compatibility constraint.

The goal of this paper, however, is not to introduce the ISE Method but rather to contextualize it within a general framework for understanding our ability to topologically redescribe our spacetime theories. Namely, this paper will introduce a representation theory for spacetime theories which is analogous to the mathematician's representation theory for groups. Recall that group representation theory helps us to systematically explore all of the possible ways in which a given group structure (e.g.,  $G\cong\text{SO}(3)$) can be realized as different sets of linear maps, $\{g:V\to V\}$, on different vector spaces, $V$. For instance, we might realize this group theoretic structure as some collection of maps, $\{g:V_5\to V_5\}\cong G$, where $\text{dim}(V_5)=5$. Alternatively, however, we might realize the same structure as some other collection of maps, $\{g:V_7\to V_7\}\cong G$, where $\text{dim}(V_7)=7$. Similarly, spacetime representation theory will help us to systematically explore all of the possible ways in which a certain pre-spacetime theoretic structure can be realized as various kinds of states on various kinds of topological stages. 



The rest of this paper is organized as follows: Sec.~\ref{SecFirstTopRed} will introduce our first example of topological redescription, namely the M\"{o}bius-Euclid duality. Sec.~\ref{SecMoreTopRed} will then work towards definitions for the terms ``pre-spacetime theory'' and ``spacetime theory'' by introducing several more examples of topological redescription. With these definitions in place, Sec.~\ref{SecStateISEThm} will then lay the groundwork for a ``spacetime representation theory'' which is in close analogy with the mathematician's group representation theory. At this point I will be able to state the ISE-Equivalence Theorem (for its proof see \citeauthor{GrimmerThesis} \citeyear[Ch.11]{GrimmerThesis}). Finally, Sec.~\ref{SecConclusion} will anticipate some of the philosophical benefits which can be expected to follow from our newfound capacity for topological redescription.

\section{A First Example of Topological Redescription: The M\"{o}bius-Euclid Duality}\label{SecFirstTopRed}
\begin{figure}[t]
\centering 
\includegraphics[width=0.475\textwidth]{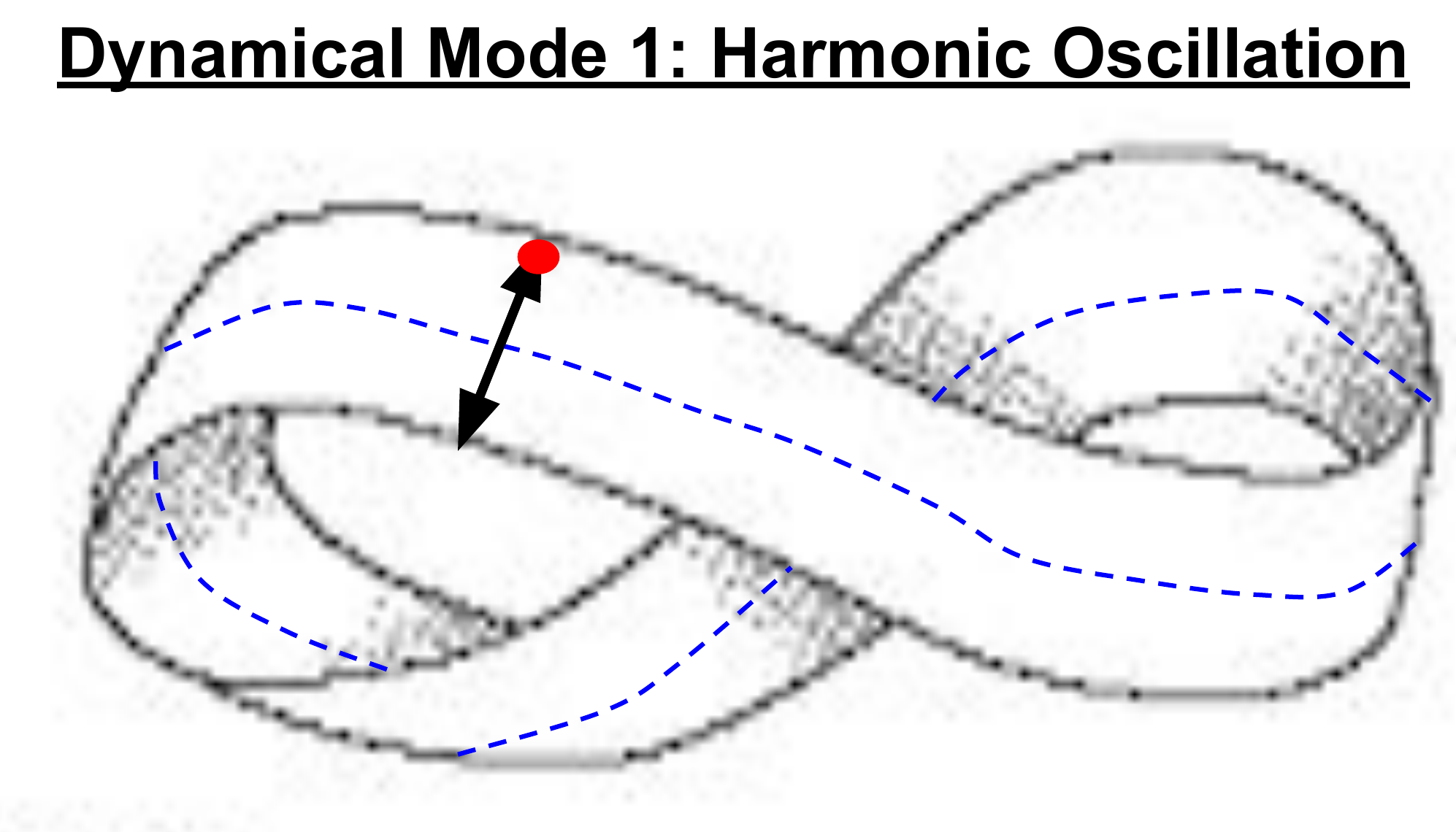}
\includegraphics[width=0.475\textwidth]{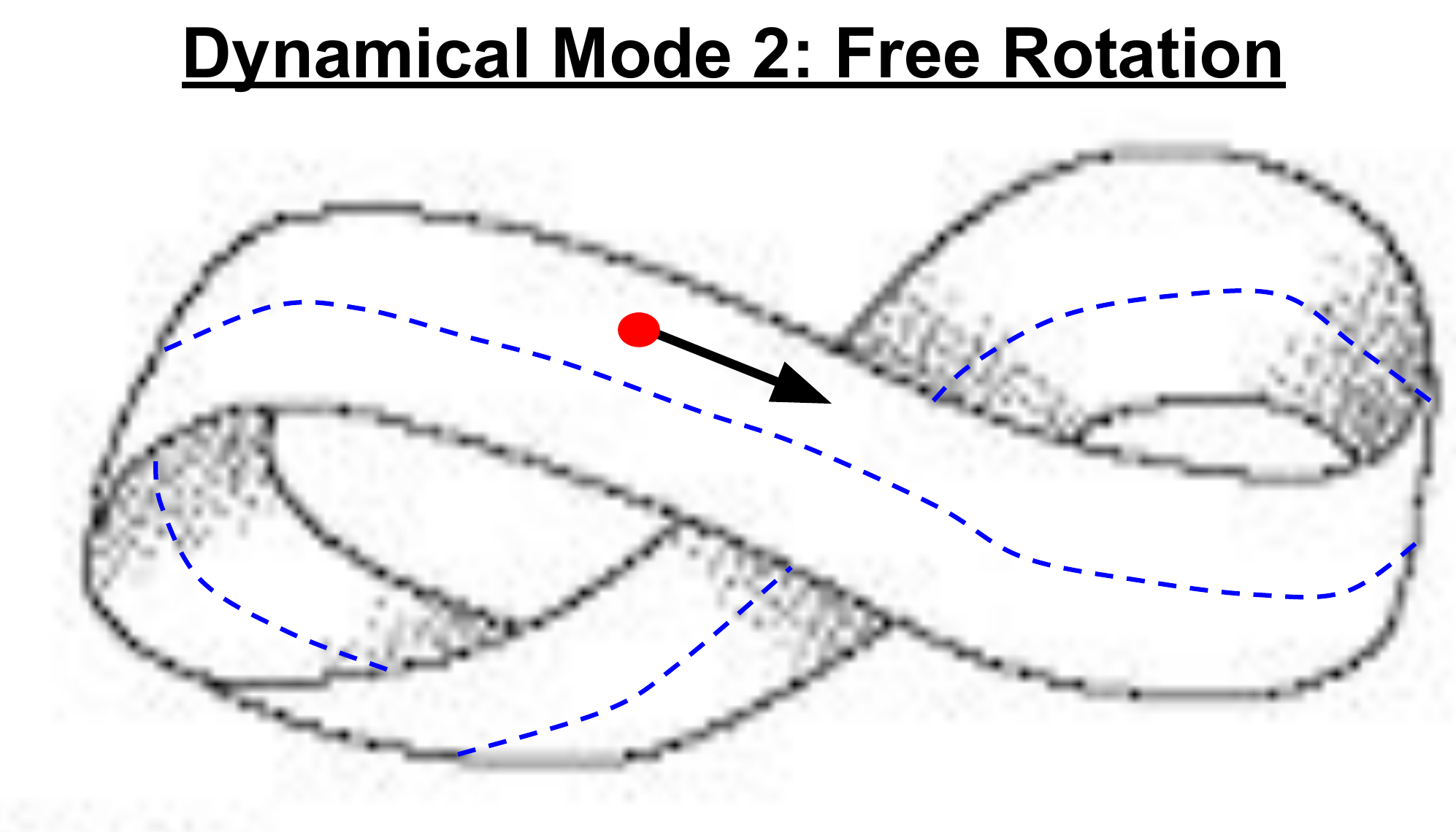}\\
\includegraphics[width=0.475\textwidth]{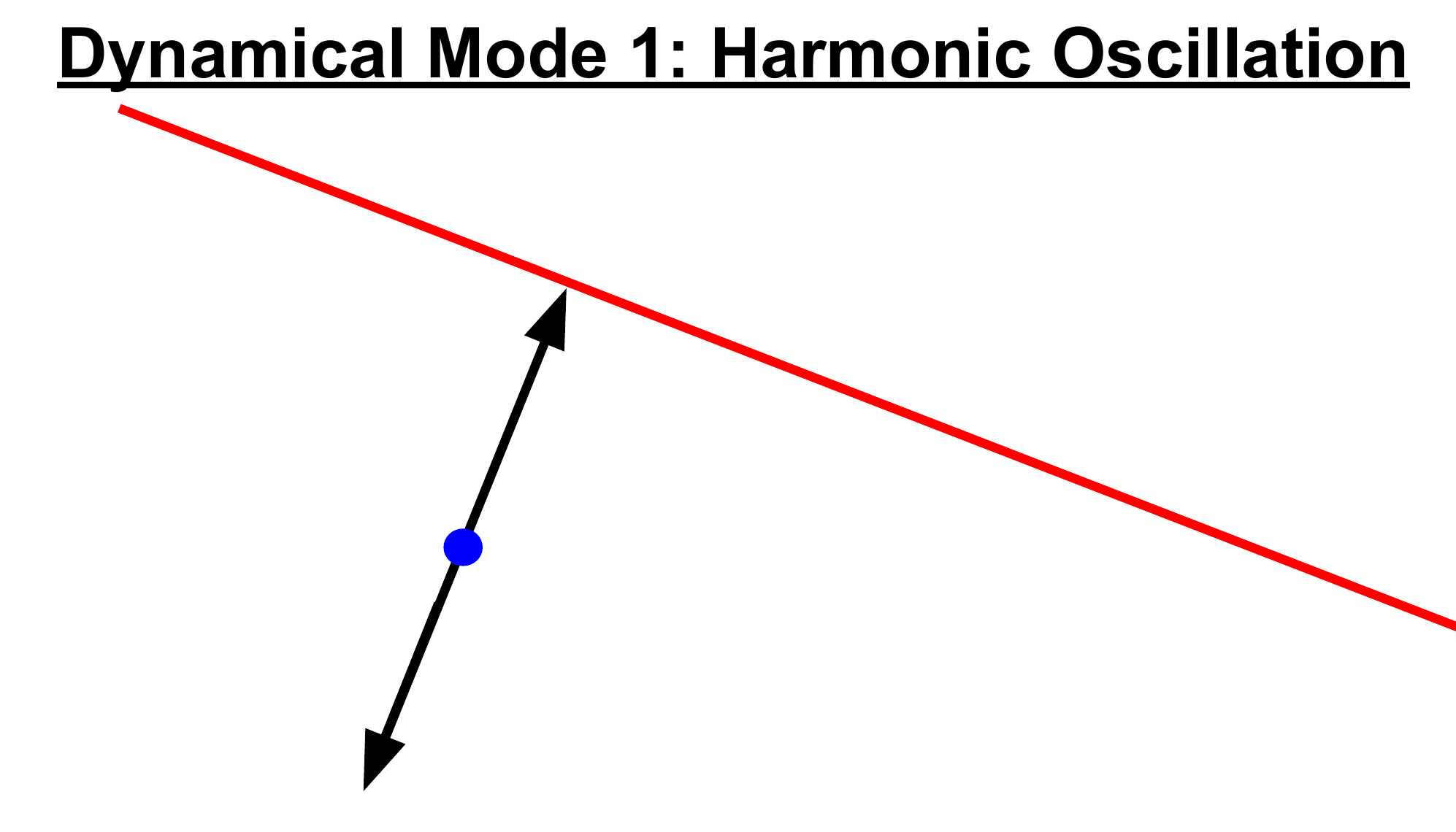}
\includegraphics[width=0.475\textwidth]{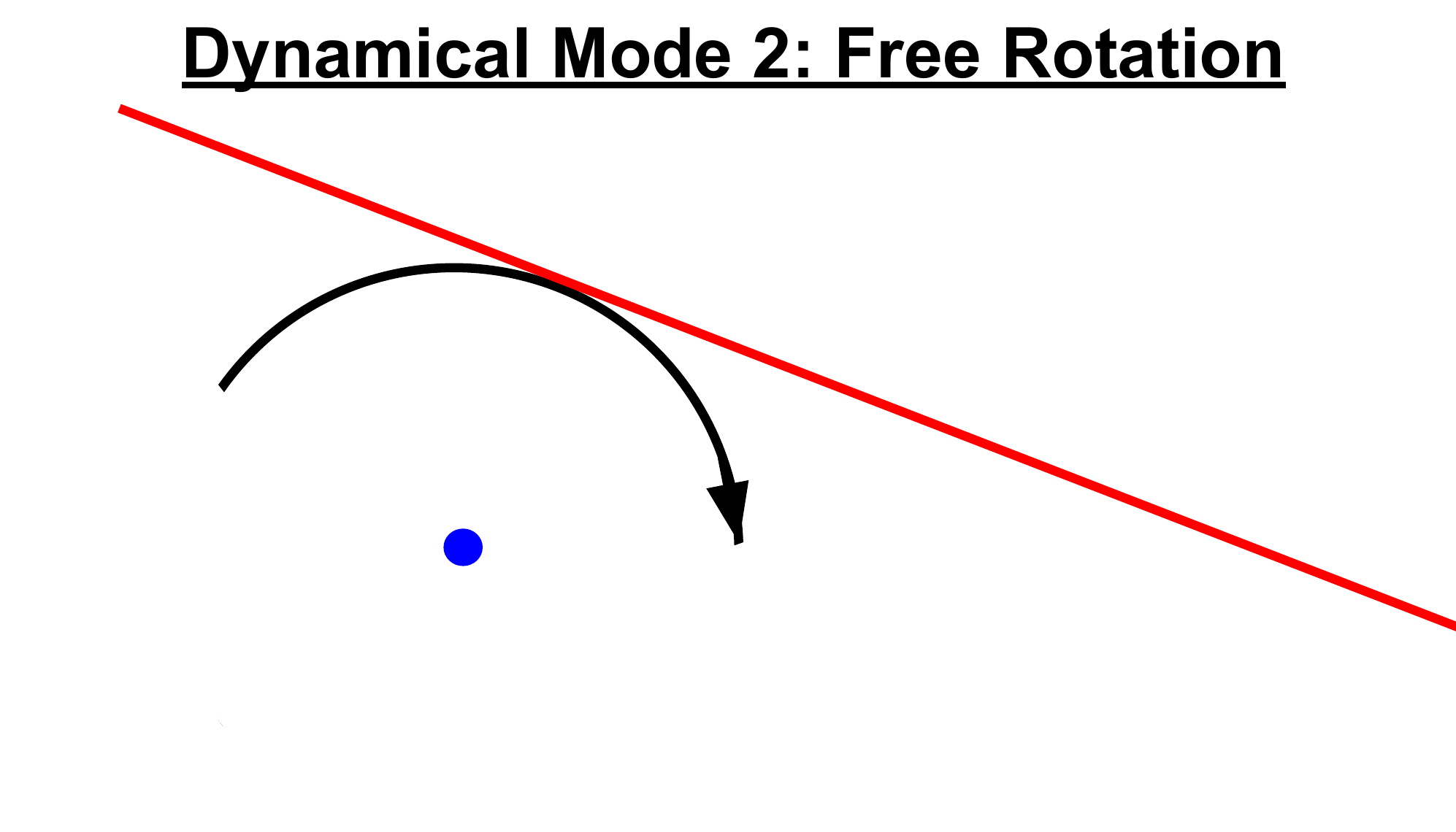}
\caption{This figure shows the dynamics of the M\"{o}bius and Euclid theories. The two subfigures on the top row depict a point-like particle (the red point) moving on a M\"{o}bius strip over time. There is a special section of the M\"{o}bius strip (the blue dashed line) about which the particle oscillates harmonically as it moves freely in the other direction. The two subfigures on the bottom row depict a line-shaped particle (the red line) moving on the Euclidean plane over time. There is a special point (the blue point) about which the particle oscillates harmonically as it rotates freely in the other direction. Perhaps surprisingly, despite having clear differences in the topological staging of their states, these theories are nonetheless equivalent in a robust sense. Namely, they are isomorphic at the level of kinematics in a way which preserves their dynamics. This is what makes them topological redescriptions of each other. Moreover, as I will discuss in Sec.\ref{SecStateISEThm}, we can think of these two theories as being different spacetime representations of the same pre-spacetime theory. \textit{(Reproduced from \cite{GrimmerThesis} with permission.)}}\label{FigMobEucDyn}
\end{figure}

\quad \ \, As a first example of topological redescription, consider the following two spacetime theories. The first theory is about a point-like particle moving on a M\"{o}bius strip, $\mathcal{M}_\text{M}\cong\mathbb{M}$, over time. Before formalizing this theory, let us first discuss its dynamics informally, see the top row of Fig.~\ref{FigMobEucDyn}. The point-like particle (in red) has two dynamical modes: 1) Neglecting any motion around the strip, the particle would undergo harmonic oscillations across the dashed blue line as if trapped in a quadratic potential. 2) Sitting at the minimum of this potential, the particle would be free to undergo uniform motion around the strip, eventually returning to its original location. Let us call the period of this motion $x=\pi$. Note that if the particle were held at a fixed distance away from the dashed blue line, then the period of this motion would instead be $x=2\pi$. That is, the particle would then have to travel twice around the M\"{o}bius strip to return to where it started. Combining these two types of motion, one can imagine the particle oscillating up and down as it travels around the M\"{o}bius strip.

The next spacetime theory which we will consider is (perhaps surprisingly) very closely related to the previous one. The key differences are as follows. Whereas the previous theory was about a particle moving on a M\"{o}bius strip, $\mathcal{M}_\text{M}\cong\mathbb{M}$, the next theory considers a particle moving on a Euclidean plane, $\mathcal{M}_\text{E}\cong\mathbb{R}^2$. Moreover, whereas the previous theory considered a point-like particle, the next theory considers an extended \textit{line-shaped particle}. (That is, in spacetime this particle would be picked out by a worldsheet rather than a worldline.) Let us again start by getting an informal grasp of this theory's dynamics, see the bottom row of Fig.~\ref{FigMobEucDyn}. As before, the particle under consideration (in red) has two dynamical modes: 1) Neglecting any rotation, the particle would undergo rigid harmonic oscillations across the special blue point as if trapped in a quadratic potential. 2) Sitting at the minimum of this potential, the particle would be free to undergo uniform rotational motion, eventually returning to its original location. Note that the period of this motion is $\theta=\pi$ because the line is unoriented . Moreover, notice that if the particle were to be held at a fixed distance away from the special blue point, then the period of this motion would instead be $\theta=2\pi$. Combining these two types of motion, one can imagine the line-shaped particle oscillating back and forth as it rigidly rotates around in the Euclidean plane.

As strange as this second spacetime theory may seem, the previous paragraph ought to sound very familiar. In fact, it is nearly identical to the above description of the M\"{o}bius theory despite the fact that these two theories have radically different topological underpinnings ($\mathcal{M}_\text{M}\cong \mathbb{M}$ vs $\mathcal{M}_\text{E}\cong \mathbb{R}^2$) and are about radically different kinds of particles (point-like vs line-shaped). Indeed, these two theories disagree about seemingly fundamental metaphysical facts: e.g., whether or not the particles in question are mereologically simple. Despite their significant differences, I will demonstrate below that (once formalized) these two theories are nonetheless kinematically and dynamically equivalent. Indeed, shockingly, there is a robust sense in which these two theories are also \textit{geometrically} equivalent, despite being topologically inequivalent.

Let us now formalize these two theories. The topological stage for the first theory includes only space and will be modeled as a M\"{o}bius strip, $\mathcal{M}_\text{M}\cong\mathbb{M}$. Since the particle described by this theory is point-like, its trajectory will be given by a function $X_\text{M}:\mathbb{R}\to\mathcal{M}_\text{M}$. If we were to have included time in $\mathcal{M}_\text{M}$, then $X_\text{M}$ would have picked out the particle's worldline. Before stating the dynamics of this particle, it will be convenient to first introduce a fixed global coordinate system for its spatial stage, $\mathcal{M}_\text{M}\cong\mathbb{M}$. For this theory, there exists a convenient coordinate map, $C:\mathbb{R}^2\to\mathcal{M}_\text{M}$, which assigns to each point, $p\in\mathcal{M}_\text{M}$, a set of coordinates,
\begin{align}\label{MobCoor}
C^{-1}(p)=\{(x+n\,\pi,(-1)^n\,y)\}_{n\in\mathbb{Z}}\subset\mathbb{R}^2.
\end{align}
One can think of advancing the $x$-coordinate as carrying us around the base space of the M\"{o}bius strip with a period of $\pi$. Advancing along the $y$-coordinate can be thought of as carrying us along the fibers of the M\"{o}bius strip. Note that the $y$-coordinate goes to infinity in both directions. In Fig.~\ref{FigMobEucDyn}, the $y$-coordinate has been cropped for visualizability.
Given this coordinate system, we can now officially state the M\"{o}bius theory as follows:
\begin{quote}
\singlespacing\vspace{-0.25cm}
\hypertarget{PartMob}{{\bf The M\"{o}bius Theory -}} Consider a theory set on a smooth manifold of $\mathcal{M}_\text{M}\cong\mathbb{M}$ about the trajectory of a particle, $X_\text{M}:\mathcal{V}_\text{M}\to\mathcal{M}_\text{M}$, with a parameter space of $\mathcal{V}_\text{M}\cong\mathbb{R}$. The only kinematic constraint on this theory's states is that $X_\text{M}$ is a smooth map. For this theory there exists a map, $C:\mathbb{R}^2\to\mathcal{M}_\text{M}$, which assigns to each point, $p\in\mathcal{M}_\text{M}$, a set of coordinates as in Eq.~\eqref{MobCoor}. In this coordinate system, the dynamical equation for the particle's trajectory is as follows:
\begin{align}\label{EqMobDyn}
\partial_\tau^2\,x(\tau) = 0,\qquad
\partial_\tau^2\,y(\tau) = -y(\tau).
\end{align}
\end{quote}
To connect with Fig.~\ref{FigMobEucDyn}, the $y=0$ coordinate on each fiber picks out the special dashed blue line in Fig.~\ref{FigMobEucDyn}.

Let us now officially introduce the Euclid theory. Whereas the previous theory considered a point-like particle, $X_\text{M}:\mathbb{R}\to\mathcal{M}_\text{M}$, the next theory considers an extended line-shaped particle, \mbox{$X_\text{E}:\mathbb{R}^2\to\mathcal{M}_\text{E}$}. If we would have included time in $\mathcal{M}_\text{E}$, then $X_\text{E}$ would have picked out the particle's worldsheet. Before I formalize this theory's dynamics, the following fact should be recalled: The equation for any straight line in the $(q_1,q_2)$-plane can be written in the form $q_1\,\text{cos}(\theta)+q_2\,\text{sin}(\theta)=r$ for some parameters $\theta$ and $r$.
\begin{quote}
\singlespacing\vspace{-0.25cm}
\hypertarget{PartLine}{{\bf The Euclid Theory -}} Consider a theory set on a smooth manifold of $\mathcal{M}_\text{E}\cong\mathbb{R}^2$ about the trajectory of a particle, $X_\text{E}:\mathcal{V}_\text{E}\to\mathcal{M}_\text{E}$, with a parameter space of $\mathcal{V}_\text{E}\cong\mathbb{R}^2$. The theory's states are subject to the following kinematic constraint: In some fixed global coordinate system, $(q_1,q_2)$, the particle's trajectory must be parameterized such that its two coordinate functions, $q_1(\tau,s)$ and $q_2(\tau,s)$, satisfy,
\begin{align}
q_1(\tau,s) \, \text{cos}(\theta(\tau)) +  q_2(\tau,s) \, \text{sin}(\theta(\tau))
=r(\tau)
\end{align}
for some smooth functions, $r(\tau)$ and $\theta(\tau)$. That is, at each fixed $\tau$-parameter the extended particle must occupy a straight line in the $(q_1,q_2)$-plane. Here $r(\tau)$ and $\theta(\tau)$ are the polar coordinates for the point on this line which is nearest to the point $(q_1,q_2)=(0,0)$. In terms of $r(\tau)$ and $\theta(\tau)$, the dynamical equation for the particle's trajectory is as follows:
\begin{align}
\partial_\tau^2\,\theta(\tau) = 0,\qquad
\partial_\tau^2\,r(\tau) = -r(\tau).
\end{align}
Note that $r(\tau)$ is here allowed to be negative with $(\theta,-r)\sim (\theta+\pi,r)$.
\end{quote}
To connect with Fig.~\ref{FigMobEucDyn}, the point with coordinates $(q_1,q_2)=(0,0)$ picks out the special blue point in Fig.~\ref{FigMobEucDyn}.

In order to see the sense in which the M\"{o}bius and Euclid theories are equivalent, let us first set aside their dynamics and focus on a fixed $\tau$ parameter for each theory (i.e., a time-slice of each theory). As I will now discuss, there is a robust geometric equivalence between the M\"{o}bius strip, $\mathcal{M}_\text{M}\cong\mathbb{M}$, and the Euclidean plane, $\mathcal{M}_\text{E}\cong\mathbb{R}^2$, despite their clear topological differences. In particular, there is a well-known isometry between the set of all unoriented lines in the Euclidean plane and the set of all points on the M\"{o}bius strip.\footnote{The isometry which I discuss below is well-known to mathematicians. See, for instance, \cite{Beem1991}. My contribution is to build a novel spacetime duality on top of this isometry.} I have constructed the above two theories such that this isometry gives a one-to-one correspondence between their dynamics. To sketch this isometry:
\begin{enumerate}
    \singlespacing
    \item[-] Every family of parallel lines in the Euclidean plane corresponds to one fiber on the M\"{o}bius strip. In particular, any two parallel lines will correspond to two points on the M\"{o}bius strip which are on the same fiber. The distance between these parallel lines corresponds to the distance-along-their-common-fiber between their corresponding points.   
    \item[-] Intersecting lines are in different parallel families and hence correspond to points on different fibers. Their acute intersection angles correspond to the angular distances between these two fibers.\footnote{Recall that we have defined the period of $\mathbb{M}$'s base space to be $x=\pi$ rather than $2\pi$. If one measures the angular distance between fibers using the latter standard, then a factor of $2$ is needed in this claim.}
\end{enumerate}

But where is the twist?\footnote{There is another way to see where the twist comes from. I claim below that the space of all \textit{unoriented} straight lines in the Euclidean plane is isomorphic to the M\"{o}bius strip; But what about oriented straight lines? One can again associate every family of parallel lines with an $\mathbb{R}$-fiber over $S^1$. This time, however, there is no twist. That is, the space of all \textit{oriented} straight lines in the Euclidean plane is isomorphic to the cylinder, $S^1\times \mathbb{R}$. Let us parameterize the cylinder as $(\theta,r)$. We can get from here to the M\"{o}bius strip by topologically identifying every line with its orientation-reversed counterpart. This amounts to taking a certain $\mathbb{Z}_2$ quotient of the cylinder, namely $(\theta,r)\sim (\theta+\pi,-r)$. One can check for oneself that $S^1\times \mathbb{R}/\sim$ is the M\"{o}bius strip, $\mathbb{M}$, as desired.} The M\"{o}bius strip famously has a topological twist to it; Where can this be seen in the above-sketched isometry? To demonstrate where the twist comes in, allow me to first state some obvious facts about the Euclidean plane before stating the isometric facts about the M\"{o}bius strip. Pick any point in the Euclidean plane to color blue (as in Fig.~\ref{FigMobEucDyn}) and consider rotating any line in the Euclidean plane around this blue point by an angle of $\theta=\pi$. Doing so, this line would return to the same parallel family that it began in, but it would have moved to the opposite side of the blue point. For most lines, you have to rotate by $\theta=\pi$ twice to get home. The exceptional cases are the lines which happen to start on the special blue point (note, the lines are unoriented). Let us now use our isometry to state some equivalent facts about the M\"{o}bius strip. Pick any point on the M\"{o}bius strip and rotate it once around the base-space (which has period $x=\pi$ by definition). Doing so, this point would end up on its original fiber but would move to the opposite side of the dashed blue line. For most points, you have to rotate by $x=\pi$ twice to get home. The exceptional cases are the points which happen to start on the blue dashed section. 

Let us now approach this isometry more formally. First note that Euclidean geometry can be formalized equally well in terms of either points or lines (i.e., by defining points in terms of the intersections of lines).\footnote{See \citeauthor{BH2016} (\citeyear[Sec.4]{BH2016}) for a discussion of this fact in the context of Morita equivalence.} Formalizing Euclidean geometry in terms of lines, it is exhausted by the set of all straight lines, $L_\text{Euc}:=\text{Lines}_\text{Euclid}$, together with their relative distances and angles. In particular, Euclidean geometry is fully captured by the quadruple, $\langle L_\text{Euc}, P_\text{Euc}, D_\text{Euc}, A_\text{Euc}\rangle$, where,
\begin{itemize}
\singlespacing
    \item[-] $P_\text{Euc}:L_\text{Euc}\times L_\text{Euc}\to\{\text{T},\text{F}\}$ reports whether or not two lines $\ell_1\in L_\text{Euc}$ and $\ell_2\in L_\text{Euc}$ are parallel to each other;
    \item[-] $D_\text{Euc}:P_\text{Euc}^{-1}(\text{T})\to\mathbb{R}_{\geq0}$ takes as input two lines which satisfy $P_\text{Euc}(\ell_1,\ell_2)=\text{T}$. That is, the two lines $\ell_1$ and $\ell_2$ are parallel. The function $D_\text{Euc}$ then reports their linear separation;
    \item[-] $A_\text{Euc}:P_\text{Euc}^{-1}(\text{F})\to[0,90^{\circ}]$ takes as input two lines which satisfy \mbox{$P_\text{Euc}(\ell_1,\ell_2)=\text{F}$}. That is, the two lines $\ell_1$ and $\ell_2$ are not parallel. The function $A_\text{Euc}$ then reports their acute angle of intersection.
\end{itemize}
These three functions, $P_\text{Euc}$, $D_\text{Euc}$, and $A_\text{Euc}$, give us a robust sense of the distance (linear or angular) between any pair of lines in the Euclidean plane.

One can construct an isomorphic quadruple in terms of the set of all points on the M\"{o}bius strip, $O_\text{M\"{o}b}:=\text{pOints}_\text{M\"{o}bius}$, as follows. Let us call two points on the M\"{o}bius strip ``parallel'' when they sit on the same fiber. We can then define the following three functions,
\begin{itemize}
\singlespacing
    \item[-] $P_\text{M\"{o}b}:O_\text{M\"{o}b}\times O_\text{M\"{o}b}\to\{\text{T},\text{F}\}$ reports whether or not two points $o_1\in O_\text{M\"{o}b}$ and $o_2\in O_\text{M\"{o}b}$ are  ``parallel'' to each other (i.e., whether or not they are on the same fiber);
    \item[-] $D_\text{M\"{o}b}:P_\text{M\"{o}b}^{-1}(\text{T})\to\mathbb{R}_{\geq0}$ takes as input two points which satisfy $P_\text{M\"{o}b}(o_1,o_2)=\text{T}$. That is, the two points $o_1$ and $o_2$ are parallel (i.e., on the same fiber). The function $D_\text{M\"{o}b}$ then reports the along-their-common-fiber distance between these two points;
    \item[-] $A_\text{M\"{o}b}:P_\text{M\"{o}b}^{-1}(\text{F})\to[0,90^{\circ}]$ takes as input two points which satisfy $P_\text{M\"{o}b}(o_1,o_2)=\text{F}$. That is, the two points  $o_1$ and $o_2$ are not parallel (i.e., they are on different fibers). The function $A_\text{M\"{o}b}$ then reports the angular separation of their fibers. Recall that we have defined the period of $\mathbb{M}$'s base space to be $x=\pi$ rather than $2\pi$. Hence, the largest possible angular separation is $\Delta x_\text{max}=\pi/2=90^{\circ}$.
\end{itemize}
These three functions, $P_\text{M\"{o}b}$, $D_\text{M\"{o}b}$, and $A_\text{M\"{o}b}$, give us a robust sense of the distance between any pair of points on the M\"{o}bius strip (measured either along or between its fibers).

Above, I claimed that the M\"{o}bius strip and the Euclidean plane are \textit{geometrically equivalent} despite their clear topological differences. Spelling this claim out formally, I meant that we have an isomorphism between the space of unoriented lines in the Euclidean plane and the space of points on the M\"{o}bius strip,
\begin{align}\label{MobEucIso}
\langle L_\text{Euc}, P_\text{Euc}, D_\text{Euc}, A_\text{Euc}\rangle\cong
\langle O_\text{M\"{o}b}, P_\text{M\"{o}b}, D_\text{M\"{o}b}, A_\text{M\"{o}b}\rangle.
\end{align}
I invite the reader to check that this isomorphism holds: The above-discussed bijection between the sets $L_\text{Euc}$ and $O_\text{M\"{o}b}$ preserves all judgements of parallel-ness, distances, and angles. This justifies calling this isomorphism an isometry. Indeed, the entirety of Euclidean geometry can be faithfully redescribed in terms of points on a M\"{o}bius strip. Any figure, any construction, and any proof which one can produce in Euclidean geometry can be equivalently demonstrated using a M\"{o}bius strip. 

Let us next use this isometry to see the sense in which the M\"{o}bius and the Euclid theories are kinematically and dynamically equivalent. To begin, note that both of these theories make a distinction between kinematically allowed and dynamically allowed states (that is, KPMs and DPMs). Let us begin by identifying the modal structure of the M\"{o}bius theory. At its broadest level, this theory is about all point-like particle trajectories on $\mathcal{M}_\text{M}\cong\mathbb{M}$ (even those which are non-smooth and/or discontinuous). That is, all $\mathbb{R}$-parameterized trajectories, $S_\text{M}^\text{all}=\{X_\text{M}:\mathcal{V}_\text{M}\to\mathcal{M}_\text{M}\}$, where $\mathcal{V}_\text{M}\cong\mathbb{R}$. Enforcing the theory's kinematic constraints, we can identify within $S_\text{M}^\text{all}$ a subset of kinematically allowed states, $S_\text{M}^\text{kin}\subset S_\text{M}^\text{all}$. This is the set of all smooth $\mathbb{R}$-parameterized trajectories on $\mathbb{M}$. Within this kinematically allowed set, we can then enforce the theory's dynamics to find its dynamically allowed states, $S_\text{M}^\text{dyn}\subset S_\text{M}^\text{kin}\subset S_\text{M}^\text{all}$.

Similarly, we can identify the modal structure of the Euclid theory as follows. At its broadest level, this theory is about all $\mathbb{R}^2$-parameterized particle trajectories on $\mathcal{M}_\text{E}\cong\mathbb{R}^2$ (even those which are non-smooth and/or discontinuous). Hence, we have that $S_\text{E}^\text{all}=\{X_\text{E}:\mathcal{V}_\text{E}\to\mathcal{M}_\text{E}\}$, where $\mathcal{V}_\text{E}\cong\mathbb{R}^2$. Enforcing the theory's kinematic constraints, we have that $S_\text{E}^\text{kin}\subset S_\text{E}^\text{all}$ is the set of all smooth $\mathbb{R}^2$-parameterized trajectories on $\mathbb{R}^2$ such that at each $\tau$-parameter $X_\text{E}(\tau,s)$ is a straight $s$-parameterized line. Within this kinematically allowed set, we can then enforce the theory's dynamics to find its dynamically allowed states, $S_\text{E}^\text{dyn}\subset S_\text{E}^\text{kin}\subset S_\text{E}^\text{all}$.

For either theory, we can think of the theory's set of kinematically allowed states, $S_\text{M/E}^\text{kin}$, as coming equipped with enough structure that we can identify within it the theory's dynamically allowed states, $S_\text{M/E}^\text{dyn}$. The details of how to do this are a bit thorny and will only be sketched below. I direct the interested reader to \citeauthor{GrimmerThesis} (\citeyear[Ch.4]{GrimmerThesis}) for complete details. Regarding the Euclid theory, we can 
write down its dynamical equation in terms of some functions on $S_\text{E}^\text{kin}$ which report the distances and angles between parallel and non-parallel lines. We have, of course, already met some such functions (namely, $P_\text{Euc}$, $D_\text{Euc}$, and $A_\text{Euc}$ defined on $L_\text{Euc}$). Let us define some analogous function on $S_\text{E}^\text{kin}$ calling them $P_\text{E}$, $D_\text{E}$, $A_\text{E}$, etc. So equipped, this theory's kinematically allowed states,
\begin{align}
\mathcal{S}_\text{E}^\text{kin}=\langle S_\text{E}^\text{kin}, P_\text{E}, D_\text{E}, A_\text{E},\dots\rangle,
\end{align}
are sufficiently structured to pick out their dynamically allowed subsets, $S_\text{E}^\text{dyn}\subset S_\text{E}^\text{kin}$. One can similarly equip the kinematically allowed states of the M\"{o}bius theory, $S_\text{M}^\text{kin}$, with some functions $P_\text{M}$, $D_\text{M}$, and $A_\text{M}$ which are analogous to the above-discussed functions $P_\text{M\"{o}b}$, $D_\text{M\"{o}b}$, and $A_\text{M\"{o}b}$ functions defined on $O_\text{M\"{o}b}$. Doing so,
\begin{align}
\mathcal{S}_\text{M}^\text{kin}=\langle S_\text{M}^\text{kin}, P_\text{M}, D_\text{M},A_\text{M},\dots\rangle,
\end{align}
is sufficiently structured to pick out this theory's dynamically allowed states, $S_\text{M}^\text{dyn}\subset S_\text{M}^\text{kin}$. 

Now that we have granted these two theories a sufficient amount of kinematic structure, we can see the sense in which they are formally equivalent to each other. Namely, the M\"{o}bius and Euclid theories are isomorphic at the level of kinematics in a way which preserves their dynamics, 
\begin{align}\label{EqKinDynEquiv}
\nonumber \text{Kinematic Isomorphism }&\text{which Preserves Dynamics:}\\
\nonumber
\langle S_\text{M}^\text{kin}, P_\text{M}, D_\text{M},A_\text{M},\dots\rangle&\cong
\langle S_\text{E}^\text{kin}, P_\text{E}, D_\text{E}, A_\text{E},\dots\rangle\\
S_\text{M}^\text{dyn}&\leftrightarrow S_\text{E}^\text{dyn}.
\end{align}
Hence, the M\"{o}bius and Euclid theories instantiate the same kinematic and dynamical structure despite differing radically in how they topologically stage their states. Indeed, as I noted above, these two theories differ not only in the topological character of their stages ($\mathcal{M}_\text{M}\cong \mathbb{M}$ vs $\mathcal{M}_\text{E}\cong \mathbb{R}^2$) but also in how they display their states upon these stages: point-like vs line-shaped, i.e., $X_\text{M}:\mathbb{R}\to\mathbb{M}$ vs $X_\text{E}:\mathbb{R}^2\to\mathbb{R}^2$. This leads us to our first pass at a definition for the term ``topological redescription'' (to be clarified at the end of Sec.~\ref{SecMoreTopRed}):
\begin{quote}
\singlespacing\vspace{-0.25cm}
{\bf Definition:} Two spacetime theories are \textit{topological redescriptions} of each other whenever they instantiate the same kinematic and dynamical structure (i.e., $\mathcal{S}_\text{new}^\text{kin}\cong \mathcal{S}_\text{old}^\text{kin}$ with $S_\text{new}^\text{dyn}\leftrightarrow S_\text{old}^\text{dyn}$) despite radically disagreeing about the topological staging of their respective states.
\end{quote}
Hence, the M\"{o}bius-Euclid duality is our first example of topological redescription.

\section{Spacetime Theories and Pre-Spacetime Theories}\label{SecMoreTopRed}
\quad \ \, This section will be spent working our way towards defining the terms ``spacetime theories'' and ``pre-spacetime theories'' in such a way that the M\"{o}bius and Euclid theories are inequivalent as spacetime theories but equivalent as pre-spacetime theories. To help demonstrate the breadth of the coming definitions, allow me to first introduce several more examples of topological redescription. As with the M\"{o}bius-Euclid duality, the following pairs of theories will exhibit the same kinematic and dynamical structure, $\mathcal{S}_\text{new}^\text{kin}\cong\mathcal{S}_\text{old}^\text{kin}$ with $S_\text{new}^\text{dyn}\leftrightarrow S_\text{old}^\text{dyn}$, despite radically disagreeing about the topological staging of their respective states.

I should first point out a subtle difference between the ``topological staging of a theory's states'' and the theory being about ``a certain kind of states on a certain topological stage''. Whereas the previous example is a clear demonstration of two theories which differ in the second way, it is actually the first notion which is at work in my definition of topological redescription. This distinction will be formalized at the end of this section, but first sketching a brief example will be helpful. 

Consider two theories which are set on isomorphic topological stages, namely, $\mathcal{M}_\text{old}\cong\mathcal{M}_\text{new}\cong\mathbb{R}^3$. Moreover, suppose that the old and new theories are both about the same kind of particle. Namely, they are both about the dynamics of a non-self-intersecting closed loop, $X_\text{old}(\tau,\theta)$ and $X_\text{new}(t,\phi)$ with $\tau,t\in\mathbb{R}$ parameterizing time and $\theta,\phi\in S^1$ parameterizing the closed loop. Finally, suppose that the two theories in question are kinematically and dynamically equivalent, $\mathcal{S}_\text{new}^\text{kin}\cong\mathcal{S}_\text{old}^\text{kin}$ with $S_\text{new}^\text{dyn}\leftrightarrow S_\text{old}^\text{dyn}$. Hence, these two theories instantiate the same kinematic and dynamical structure by using the same kind of states on the same kind of stage. Despite all of their similarities, these two theories might nonetheless differ in the topological staging of their states.

But how is this possible? The old theory's states might, for instance, be topologically knotted in a specific way whereas the new theory's states are unknotted. While the two theory's stages are isomorphic and their states are of the same kind, there is nonetheless a clear topological difference regarding how their states fit on this stage; There is no topology preserving way to map the old theory's states onto the new theory's states, nor vice versa. In particular, the assumed-to-exist map, $S_\text{new}^\text{kin}\leftrightarrow S_\text{old}^\text{kin}$, which relates the two theories' kinematic structures cannot be the lift of a diffeomorphism, $\mathcal{M}_\text{new}\leftrightarrow \mathcal{M}_\text{old}$, which relates their stages. In sum, the ``topological staging of a theory's states'' includes facts about the theory's stage and the theory's states as well as facts about the particular \textit{staging} of the latter on the former. 

With this distinction in mind, let us now spell out in detail our second example of topological redescription. Consider the following spacetime theory about a modified quantum harmonic oscillator (QHO):\footnote{This duality between the Non-Local QHO and the Cosine QHO theories was first discussed in \cite{RasmusEnrico}.}
\begin{quote}
\singlespacing\vspace{-0.25cm}
\hypertarget{NLQHO}{{\bf Non-Local QHO -}} Consider a theory set on a smooth manifold, \mbox{$\mathcal{M}_\text{old}\cong\mathbb{R}^2$}, about a scalar field, $\varphi_\text{old}:\mathcal{M}_\text{old}\to\mathcal{V}_\text{old}$, with a value space of $\mathcal{V}_\text{old}\cong\mathbb{C}$. This theory's states are subject to the following kinematic constraint: In some fixed global coordinate system, $(t,x)$, at each fixed $t$-coordinate the state must be within the usual rigged Hilbert space of planewaves and Dirac deltas. At each $t$-coordinate, the metaphysically relevant way to judge the relative size and similarity of this theory's states is with an $L^2$ inner product. In this coordinate system, the dynamical equation for the field is as follows:
\begin{align}
\ii\partial_t\varphi_\text{old}(t,x)
&=(-\partial_x^2+x^2)\varphi_\text{old}(t,x)+ \frac{\lambda}{2} \varphi_\text{old}(t,x-a)+ \frac{\lambda}{2} \varphi_\text{old}(t,x+a),
\end{align}
for some fixed $a>0$ and $\lambda\in\mathbb{R}$.
\end{quote}
It should be noted that by using the identity, $h(x+a)=\exp(a\partial_x) h(x)$, this theory's dynamical equation can be rewritten as:
\begin{align}
\ii\partial_t\varphi_\text{old}(t,x)
&=(-\partial_x^2+x^2+\lambda\text{cosh}(a\partial_x))\varphi_\text{old}(t,x).
\end{align}
We can recognize this dynamical equation as the Schr\"{o}dinger equation with a (rather bizarre) Hamiltonian of $H_\text{old}=-\partial_x^2+x^2+\lambda\,\text{cosh}(a\partial_x)$. This theory describes the dynamics of a standard QHO plus a non-local self-coupling. 

To complete our second example of topological redescription, consider also the following closely related spacetime theory about a differently modified quantum harmonic oscillator (QHO):
\begin{quote}
\singlespacing\vspace{-0.25cm}
\hypertarget{CosQHO}{{\bf Cosine QHO -}} Consider a theory set on a smooth manifold, \mbox{$\mathcal{M}_\text{new}\cong\mathbb{R}^2$}, about a scalar field, $\varphi_\text{new}:\mathcal{M}_\text{new}\to\mathcal{V}_\text{new}$, with a value space, $\mathcal{V}_\text{new}\cong\mathbb{C}$. The theory's states are subject to the following kinematic constraint: In some fixed global coordinate system, $(\tau,q)$, at each fixed $\tau$-coordinated the state must be within the usual rigged Hilbert space of planewaves and Dirac deltas. At each $\tau$-coordinate, the metaphysically relevant way to judge the relative size and similarity of this theory's states is with an $L^2$ inner product. In this coordinate system, the dynamical equation for the field is as follows:
\begin{align}
\ii\partial_\tau\varphi_\text{new}(\tau,q)
&=(-\partial_q^2+q^2+\lambda\,\text{cos}(a\,q))\,\varphi_\text{new}(\tau,q),
\end{align}
for some $a>0$ and $\lambda\in\mathbb{R}$.
\end{quote}
We can again recognize this as the Schr\"{o}dinger equation but this time with a Hamiltonian of $H_\text{new}=-\partial_q^2+q^2+\lambda\,\text{cos}(a\,q)$. This Hamiltonian describes the dynamics of a standard QHO plus an extra cosine-shaped potential. 

The keen reader may have noticed that the above two theories are related to each other via a Fourier transform.\footnote{In particular they are related by the replacements $x\mapsto \ii \partial_q$ and $\partial_x\mapsto \ii q$ noting that $\text{cosh}(\ii a z) = \text{cos}(a z)$.} Indeed, they are an instance of the position-momentum duality of non-relativistic quantum mechanics. Note that as in the above-sketched example, the topological stages of these two theories are diffeomorphic, $\mathcal{M}_\text{old}\cong\mathcal{M}_\text{new}\cong\mathbb{R}^2$. Importantly, however, the Fourier transform which relates these two theories does not respect this diffeomorphism; Indeed, it radically reshapes how the theory's states sit on their respective spacetime manifolds. The old theory's planewaves become Dirac deltas in the new theory and vice versa. More generally, what was non-local becomes local and vice-versa. This is evidenced by the fact that the old theory's non-local dynamics has become local dynamics in the new theory. As this example of topological redescription shows, it is possible to dramatically change the topological staging of a theory's states while preserving the topological character of their respective stages, $\mathcal{M}_\text{old}\cong\mathcal{M}_\text{new}$. 

Despite the stark differences which these two QHO theories have in the topological staging of their states and dynamics, there is nonetheless a robust sense in which these theories are structurally equivalent. Namely, as I will now discuss, we can see them as being kinematically and dynamically equivalent \textit{in the exact same sense} as the M\"{o}bius and Euclid theories are. Namely, these two theories are isomorphic at the level of kinematics, and this isomorphism preserves the theories' dynamics. To see this explicitly, we will need to first identify the modal structure of the above two theories (i.e., their DPMs and KPMs, $S_\text{old}^\text{dyn}\subset S_\text{old}^\text{kin}$ and $S_\text{new}^\text{dyn}\subset S_\text{new}^\text{kin}$). We will then need to equip $\mathcal{S}_\text{old}^\text{kin}=\langle S^\text{kin}_\text{old},\dots\rangle$ and $\mathcal{S}_\text{new}^\text{kin}=\langle S^\text{kin}_\text{new},\dots\rangle$ with at least enough structure so that we can identify within them their dynamically allowed states, $S_\text{old}^\text{dyn}\subset S_\text{old}^\text{kin}$ and $S_\text{new}^\text{dyn}\subset S_\text{new}^\text{kin}$.

Let us begin by identifying the modal structure of the Non-Local QHO theory. In the broadest sense, this theory is about the set, $S^\text{all}_\text{old}$, of all $\mathbb{C}$-valued functions definable on $\mathcal{M}_\text{old}\cong\mathbb{R}^2$ respectively (even those which are non-smooth and/or discontinuous). Enforcing the theory's kinematic constraints, we can identify within $S^\text{all}_\text{old}$ a set of kinematically allowed states, $S^\text{kin}_\text{old}\subset S^\text{all}_\text{old}$. This is the set of all $\mathbb{C}$-valued functions within ``the usual rigged Hilbert space of planewaves and Dirac deltas''. Within this set, we can then enforce the theory's dynamics to find its dynamically allowed states, $S^\text{dyn}_\text{old}\subset S^\text{kin}_\text{old}\subset S^\text{all}_\text{old}$. We can similarly identify the modal structure of the Cosine QHO theory, $S^\text{dyn}_\text{new}\subset S^\text{kin}_\text{new}\subset S^\text{all}_\text{new}$, by applying the same analysis except with $\mathcal{M}_\text{new}\cong\mathbb{R}^2$ replacing $\mathcal{M}_\text{old}\cong\mathbb{R}^2$. For notational convenience, I will now drop the new/old subscripts whenever either could apply.

Let us next equip these two theories with enough kinematic structure to state their dynamical equations. Given that the relevant dynamical equations are both linear, all that we need to do is to equip $S^\text{kin}_\text{new}$ and $S^\text{kin}_\text{new}$ the structure of a vector space. Namely, we can give both of these sets an addition operation, $+:S^\text{kin}\times S^\text{kin}\to S^\text{kin}$, and a scalar multiplication operation, $\cdot:\mathbb{C}\times S^\text{kin}\to S^\text{kin}$. If the dynamics of these theories had been non-linear, then we would have additionally required a product operation, $\times:S^\text{kin}\times S^\text{kin}\to S^\text{kin}$. Using just these structures we can state the theory's dynamical equations and define its dynamically allowed states, $S^\text{dyn}\subset S^\text{kin}$. 

It is important to note that a theory's kinematic structures might go beyond the bare minimum which are required to state its dynamical equations. For instance, note the empirical role that Hilbert space structure plays in non-relativistic quantum mechanics despite the fact that no inner product appears in the Schr\"{o}dingder equation. Relatedly, this inner product plays a significant role in our ongoing efforts to interpret quantum mechanics. Hence, for either empirical or metaphysical reasons, our two QHO theories should come equipped with an inner product in addition to their vector space structure. More generally, our spacetime theories might come equipped with some structures which have been flagged as being empirically and/or metaphysically relevant. In my coming definition of the term ``spacetime theory'', I shall demand that our spacetime theories come equipped with all such structures. (I will soon call this the kinematic structure assumption.)

Of course, deciding which aspects of our theories are empirically, and/or metaphysically relevant will require that we go beyond a merely formal analysis of our theories. The kinematic structure assumption hence means that---before we can begin topologically redescribing a given theory---it must be at least partly interpreted, e.g., connected up with our experimental and inferential practices. The exact degree of pre-interpretation which is required, however, will depend upon one's philosophical goals. More will be said on this point in Sec.~\ref{SecConclusion}. For now, however, let us continue in a purely formal mode.

As I noted above, our two QHO theories should come equipped with an inner product which has been formally flagged as empirically and/or metaphysically relevant. To define this inner product let us introduce an auxiliary space, $T_\text{old}\cong T_\text{new}\cong\mathbb{R}$, which parameterizes time for each theory. Using this space we can define at each parameter, $s\in T$, an inner product,
\begin{align}\label{InnerProduct}
I(\varphi,\phi,s)\coloneqq\langle\varphi,\phi\rangle(s)\coloneqq\int_{-\infty}^\infty \varphi(s,y)^* \phi(s,y) \, \d y.
\end{align}
Namely, for either the new or old theory we have a function $I:S^\text{kin}\times S^\text{kin}\times T\to \mathbb{C}$ for judging the relative size and similarity of any two kinematically allowed states, $\varphi,\phi\in S^\text{kin}$ at a parameter, $s\in T$.

In total, for the two QHO theories we have identified the following kinematic structures as being dynamically, empirically, and/or metaphysically relevant,
\begin{align}\label{OldQHOFormal0}
\mathcal{S}_\text{old}^\text{kin}&\coloneqq \langle S_\text{old}^\text{kin},+_\text{old},\,\cdot_\text{old}\, , T_\text{old},\,I_\text{old}\,\rangle,\\
\label{NewQHOFormal0}
\mathcal{S}_\text{new}^\text{kin}&\coloneqq \langle S_\text{new}^\text{kin},+_\text{new},\,\cdot_\text{new}\, , T_\text{new},\,I_\text{new}\rangle.
\end{align}
That is, at a kinematic level, both of these theories have the structure of an inner product space (at each parameter $s\in T$). In fact (glossing over some technical details), both of these theories have the structure of a Hilbert space at each parameter $s\in T$. For both theories, this Hilbert space structure is sufficient to define the theory's dynamically allowed states, $S^\text{dyn}\subset S^\text{kin}$. As with the M\"{o}bius and Euclid theories, there is a clear sense in which these two QHO theories are kinematically and dynamically equivalent.
\begin{align}
\nonumber \text{Kinematic Isomorphism }&\text{which Preserves Dynamics:}\\
\nonumber\langle S_\text{old}^\text{kin},+_\text{old},\,\cdot_\text{old}\, , T_\text{old},\,I_\text{old}\,\rangle &\cong \langle S_\text{new}^\text{kin},+_\text{new},\,\cdot_\text{new}\, , T_\text{new},\,I_\text{new}\rangle\\
S_\text{old}^\text{dyn}&\leftrightarrow S_\text{new}^\text{dyn}.
\end{align}
Namely, despite their differences in topological staging, we nonetheless have $\mathcal{S}_\text{old}^\text{kin}\cong \mathcal{S}_\text{new}^\text{kin}$ with $S_\text{old}^\text{dyn}\leftrightarrow S_\text{new}^\text{dyn}$. Hence, these two QHO theories qualify as our second example of topological redescription.

Before officially defining the terms ``spacetime theory'' and ``pre-spacetime theory'', allow me to now (more quickly) introduce two more examples of topological redescription, which I have  discussed at length in \citeauthor{GrimmerThesis} (\citeyear[Ch.8]{GrimmerThesis}). Both of these examples start with a theory about a $\mathbb{R}$-valued scalar field, \mbox{$\varphi_\text{old}:\mathcal{M}_\text{old}\to \mathbb{R}$} on some spacetime manifold, $\mathcal{M}_\text{old}$. In each case, one can topologically redescribe the theory in question as being about some other field, \mbox{$\varphi_\text{new}:\mathcal{M}_\text{new}\to \mathcal{V}_\text{new}$}, valued elsewhere, $\mathcal{M}_\text{new}$, and potentially valued differently, $\mathcal{V}_\text{new}$. In the first case, one can begin from a lattice theory (i.e., a theory set on a discrete spacetime, $\mathcal{M}_\text{old}\cong\mathbb{R}\times\mathbb{Z}$) and topologically redescribe it as taking place on a continuous spacetime manifold, $\mathcal{M}_\text{new}\cong\mathbb{R}\times\mathbb{R}$.\footnote{It was through attempting to generalize an example of this kind that I discovered the ISE Method. This inspiring example was itself inspired by the work of the brilliant mathematical physicist \cite{Kempf_2010} entitled, ``Spacetime could be simultaneously continuous and discrete, in the same way that information can be''. For an overview of Kempf's work on this topic, see~\cite{Kempf2018}. In fact, it was Kempf's work which first motivated me to look into the philosophical implications of topological redescription.} 

In the other case, one can begin with a theory with some compactified dimensions, $\mathcal{M}_\text{old}\cong\mathbb{R}^2\times S^2$, and end up on a lower dimensional spacetime, $\mathcal{M}_\text{new}\cong\mathbb{R}^2$. The cost, however, is that our old $\mathbb{R}$-valued field, $\varphi_\text{old}:\mathcal{M}_\text{old}\to \mathbb{R}$, will become an $\mathbb{R}^\infty$-valued field, $\varphi_\text{new}:\mathcal{M}_\text{new}\to \mathbb{R}^\infty$, whose components, $\varphi_\text{new}^{(\ell m)}:\mathcal{M}_\text{new}\to\mathbb{R}$, correspond to the harmonic mode of the old theory's compactified dimensions.

Without further ado, let us now define the terms ``pre-spacetime theory'' and ``spacetime theory'' with enough generality to capture all of the above examples. Let us first try to capture the sense in which the M\"{o}bius and Euclid theories (as well as our other examples) are nonetheless inequivalent, despite their similarities. Indeed, per the following definition, they are inequivalent \textit{as spacetime theories}.
\begin{quote}
\singlespacing\vspace{-0.25cm}
{\bf Definition:} \textit{Spacetime Theory} - In order to qualify as a spacetime theory, the theory under consideration must satisfy the following three assumptions:
\begin{enumerate}
    \item[1.] \textit{Modal Assumption:} The theory must distinguish between dynamically allowed states and kinematically allowed states as $S^\text{dyn}\subset S^\text{kin}$. That is, it must have DPMs and KPMs. 
    \item[2.] \textit{Kinematic Structure Assumption:} At the level of kinematics, the theory must come equipped with all of its dynamically, empirically, and metaphysically relevant structures, $\mathcal{S}^\text{kin}=\langle S^\text{kin},\dots\rangle$. These structures must allow us to identify the theory's dynamically allowed states as $S^\text{dyn}$ as a subset of $S^\text{kin}$.
    \item[3.] \textit{Declared Topological Staging Assumption:}\footnote{It should be noted that---per this definition---general relativity (GR) is not a spacetime theory; Instead, one can think of it as a schema for spacetime theories. For instance, GR with $\mathcal{M}\cong\mathbb{R}^4$ is a spacetime theory. So too is GR with $\mathcal{M}\cong\mathbb{R}^2\times S^2$.} The theory must be formalized with a declared topological staging for its states. It must specify a specific stage for its states, $\Sigma$, and declare that it is about a certain kind of states, $S^\text{all}\supset S^\text{kin}\supset S^\text{dyn}$ defined on $\Sigma$. Moreover, there must be a canonical way to lift $\Sigma$'s automorphisms, $d:\Sigma\to\Sigma$, to act on the theory's states, $d^*:S^\text{all}\to S^\text{all}$.
    
    More generally, we might have two copies of $\Sigma$ and thereby two copies of $S^\text{all}$. We should then be able to lift any isomorphism $d:\Sigma_\text{old}\to\Sigma_\text{new}$, to act as a bijection on the theories' states as $d^*:S_\text{old}^\text{all}\to S_\text{new}^\text{all}$. Moreover, this $*_\text{lift}:d\mapsto d^*$ map must be a group homomorphism, i.e., we must have $d_1^*\,d_2^*=(d_1\,d_2)^*$.
\end{enumerate}
\end{quote}
The reader is invited to check that all of these assumptions are met by the M\"{o}bius theory, the Euclid theory, and both of our QHO theories.

Regarding the kinematic structure assumption, as the above-discussed examples have hopefully demonstrated, the types of kinematic structures, $\mathcal{S}^\text{kin}\coloneqq \langle S^\text{kin}, \dots \rangle$, which we associate with our theories can vary significantly from case to case. For the M\"{o}bius and Euclid theories discussed above, their kinematic structures had a notable geometric flavor to them (``parallel'', ``distances'', ``angles'', etc.). By contrast, for the two QHO theories these structures had a more algebraic flavor to them (``addition'', ``scalar multiplication'', ``inner products'', etc.). In general, the theory's kinematically allowed states, $\mathcal{S}^\text{kin}$, are allowed to be equipped with effectively any kind of structures one wishes.

Regarding the declared topological staging assumption, we have a wide range of possibilities available to us. For instance, if the theory declares itself to be set on a smooth manifold, $\Sigma=\mathcal{M}$, then it could be about a particle trajectory, $X:\mathcal{V}\to\mathcal{M}$, or a scalar field, $\varphi:\mathcal{M}\to\mathcal{V}$, or a tangent vector field, $\vec{\varphi}:\mathcal{M}\to\text{T}\mathcal{M}$. Alternatively, the theory might be about the global sections, $\varphi:B\to E$, of some fiber bundle, $\Sigma=\langle E,B,F,\pi:E\to B\rangle$. More exotically, the theory could be set on a supermanifold (\`{a} la supersymmetry) or on a topological space (e.g., a non-Hausdorff manifold). In any case, the above-discussed pairs of theories are ultimately inequivalent as spacetime theories because of their differences in the topological staging of their states.

As I noted above, there is more to the ``topological staging of a theory's states'' than it being about a certain kind of state on a certain kind of stage. Indeed, as I will discuss momentarily, defining a sense of (in-)equivalence between spacetime theories requires the lifting map, $*_\text{lift}:d\mapsto d^*$, which connects an action on the theory's stage, $d$, to an action on its states, $d^*$. Before seeing this, however, let me quickly confirm that this lifting map exists for a wide range of theories. For theories set on a smooth manifold $\Sigma=\mathcal{M}$ about a scalar field, $\varphi:\mathcal{M}\to \mathcal{V}$, we have $d^*\varphi\coloneqq \varphi\circ d^{-1}$. Alternatively, for theories about a particle trajectory, $X:\mathcal{V}\to\mathcal{M}$, we can simply take  $d^*X\coloneqq d\circ X$. Our theory could even be about a tangent vector field, $\vec{\varphi}:\mathcal{M}\to \text{T}\mathcal{M}$, in which case we would have that $d\in\text{Diff}(\mathcal{M})$ lifts to act on $\vec{\varphi}$ as $d^*\vec{\varphi}$ via the push-forward. If we were to restrict our attention to cases where $\Sigma=\mathcal{M}$, the lifting assumption demands that the theory under consideration be ``natural'' in the sense of \cite{NaturalGaugeNatural}. 

It is important to note, however, that the lifting assumption is also met by generic theories about sections $\varphi:B\to E$ of a fiber bundle $\Sigma=\langle E,B,F,\pi:E\to B\rangle$. The relevant automorphisms $d\in \text{Auto}(E)\subset\text{Diff}(E)$ are those diffeomorphisms on $E$ which map fibers to fibers. Note that these are exactly the diffeomorphisms on $E$ which have a well-defined image on the base space, $\pi[d]:B\to B$. Using this $\pi[d]$ map, we can lift any $d\in \text{Auto}(E)$ to act on any global section, $\varphi:B\to E$, as $d^*\varphi\coloneqq d\circ\varphi\circ\pi[d^{-1}]$.

Given the lifting map, $*_\text{lift}:d\mapsto d^*$, we can make the following definition.
\begin{quote}
{\bf Definition: Equivalence as Spacetime Theories} - Two theories are \textit{equivalent as spacetime theories} if there is an isomorphism between their stages $d:\Sigma_\text{old}\to\Sigma_\text{new}$ which lifts to an isomorphism at the level of kinematics, $d^*\vert_\text{kin}:\mathcal{S}_\text{old}^\text{kin}\leftrightarrow \mathcal{S}_\text{new}^\text{kin}$, which preserves the theories' dynamics, $d^*\vert_\text{dyn}:S_\text{old}^\text{dyn}\leftrightarrow S_\text{new}^\text{dyn}$.    
\end{quote}
Hence, the M\"{o}bius and Euclid theories are inequivalent as spacetime theories because their stages are non-isomorphic, $\mathcal{M}_\text{old}\not\cong\mathcal{M}_\text{new}$. For the two QHO theories, however, their stages are isomorphic, $\mathcal{M}_\text{old}\cong\mathcal{M}_\text{new}$, but they are nonetheless inequivalent since their kinematic isomorphism (i.e., the Fourier transform) is not the lift of any of the diffeomorphisms which relate $\mathcal{M}_\text{old}$ to $\mathcal{M}_\text{new}$. Two theories can differ in the topological staging of their states while being about the same kinds of states on topologically equivalent stages.

There is also a sense in which the above-discussed pairs of theories are nonetheless equivalent to each other despite their differences in topological staging. Indeed, per the following definition, they are equivalent \textit{as pre-spacetime theories}.
\begin{quote}
\vspace{-0.5cm}\singlespacing
{\bf Definition:} \textit{Pre-Spacetime Theory} - In order to qualify as a pre-spacetime theory, the theory under consideration must satisfy the following three assumptions:
\begin{enumerate}
    \item[1.] \textit{Modal Assumption:} Same as above. 
    \item[2.] \textit{Kinematic Structure Assumption:} Same as above.
    \item[3.] \textit{No Declared Topological Staging Assumption:} The theory must be formalized without a declared topological staging for its states. That is $\mathcal{S}^\text{kin}=\langle S^\text{kin},\dots\rangle$ must be formalized with  $S^\text{kin}$ being a set of ur-elements with no internal structure and \textit{not as a set of maps} e.g., $\{X:\mathcal{V}\to\mathcal{M}\}$ or $\{\varphi:\mathcal{M}\to\mathcal{V}\}$ or $\{\vec{\varphi}:\mathcal{M}\to\text{T}\mathcal{M}\}$ or $\{\varphi:B\to E\}$.
\end{enumerate}
The criteria for equivalence of two pre-spacetime theories is that they are isomorphic at the level of kinematics, $\mathcal{S}_\text{old}^\text{kin}\cong \mathcal{S}_\text{new}^\text{kin}$, in a way which preserves their dynamics, $S_\text{old}^\text{dyn}\leftrightarrow S_\text{new}^\text{dyn}$.
\end{quote}
The two QHO theories are equivalent in this way despite being inequivalent as spacetime theories. The same is true for the M\"{o}bius and Euclid theories. Indeed, it was precisely this fact about the M\"{o}bius-Euclid duality which motivated our definition of topological redescription at the end of Sec.~\ref{SecFirstTopRed}. Hence, we can now rephrase this definition as follows:
\begin{quote}
\singlespacing\vspace{-0.25cm}
{\bf Rephrased Definition:} Two spacetime theories are \textit{topological redescriptions} of each other whenever they are equivalent as pre-spacetime theories but inequivalent as spacetime theories.  
\end{quote}
We are now in a position to discuss spacetime representation theory and the ISE-Equivalence Theorem.

\section{Spacetime Representation Theory and the ISE-Equivalence Theorem}\label{SecStateISEThm}
\quad \ \, The previous section has ended with a revised definition of topological redescription. The sensibility of this definition, however, turns upon our ability to alternatively conceive of a given theory as either a spacetime theory or as a pre-spacetime theory. To demonstrate how this is possible, a comparison with the representation theory of groups will be helpful. Recall that group representation theory is about exploring the different ways in which a given group structure (e.g.,  $G=\text{SO}(3)$) can be realized as sets of linear maps, $\{g:V\to V\}$, on various vector spaces, $V$. For instance, we might have,
\begin{align}
G_5=\{g:V_5\to V_5\}\quad\text{or alternatively}\quad G_7=\{g:V_7\to V_7\},    
\end{align}
where $\text{dim}(V_5)=5$ and $\text{dim}(V_7)=7$. If we think of $G_5$ and $G_7$ as group representations (treating $g:V\to V$ as a map from $V$ to itself) then we have $G_5\not\cong G_7$ because we have $V_5\not\cong V_7$ as vector spaces. We can, however, still treat $G_5$ and $G_7$ as groups, by simply ignoring the internal structure of their elements, (i.e., by treating $g$ as an ur-element). Doing so we would find that we have $G_5\cong G_7$ as groups. In sum, $G_5$ and $G_7$ are equivalent as groups but inequivalent as group representations. Said differently, they are inequivalent representations of the same group theoretic structure.

Importantly, however, there is more to a group representation than the fact that realizes a certain group theoretic structure, $G$, as maps on a certain vector space, $V$. The criteria for two group representations $G_A=\{g:V_A\to V_A\}$ and $G_B=\{g:V_B\to V_B\}$ to be considered equivalent is as follows. There must be an isomorphism between the vector spaces, $U:V_A\to V_B$ which lifts as $U^*g\coloneqq U \, g \, U^{-1}$ to a group theoretic isomorphism between $G_A$ and $G_B$. Hence it is possible for two group representations to be inequivalent while representing the same group theoretic structure, $G_A\cong G_B$, on two isomorphic vector spaces, $V_A\cong V_B$. The reader has hopefully noticed some parallels with my above discussion of topological staging.

Indeed, this analogy motivates us to adopt the same terminology for spacetime theories and pre-spacetime theories. For instance, we can analogously think of the M\"{o}bius and Euclid theories as being different spacetime representations of the same pre-spacetime theory. The M\"{o}bius theory realized their common pre-spacetime theoretic structure as a certain kind of states, $S_\text{M}^\text{kin}=\{X_\text{M}:\mathbb{R}\to\mathbb{M}\}$. By contrast, the Euclid theory realizes this same pre-spacetime theoretic structure as a different kind of states, $S_\text{E}^\text{kin}=\{X_\text{E}:\mathbb{R}^2\to\mathbb{R}^2\}$. More subtly, the QHO example has demonstrated how two spacetime representations might be inequivalent despite being about the same kind of states on the same kind of stage. Hence, by analogy, we can think of a \textit{spacetime representation theory} which explores the different ways in which a given pre-spacetime theoretic structures, $\mathcal{S}^\text{kin}=\langle S^\text{kin},\dots\rangle$, can be realized as various kinds of states arranged in various ways on various kinds of topological stages. 

To make this more concrete, suppose that we begin with a spacetime theory (e.g., about a scalar field, $\varphi_\text{old}:\mathcal{M}_\text{old}\to\mathbb{R}$) which has some kinematic/dynamical structure, namely $\mathcal{S}_\text{old}^\text{kin}=\langle S_\text{old}^\text{kin},\dots\rangle$ and $S_\text{old}^\text{dyn}\subset S_\text{old}^\text{kin}$. Thinking of this as a pre-spacetime theory, we can then explore different spacetime representations of this theory, $\mathcal{S}_\text{new}^\text{kin}\cong \mathcal{S}_\text{old}^\text{kin}$ with $S_\text{new}^\text{dyn}\leftrightarrow S_\text{old}^\text{dyn}$. For instance, we might try to represent its states as being:
\begin{itemize}
\singlespacing
    \item[-] $\mathbb{R}$-valued scalar fields, $S_\text{new}^\text{kin}=\{\varphi:\mathcal{M}_0\to \mathbb{R}\}$, on a smooth manifold $\mathcal{M}_0\cong\mathcal{M}_\text{old}$ but \textit{arranged differently}, or 
    \item[-] $\mathbb{R}^\infty$-valued scalar fields, $S_\text{new}^\text{kin}=\{\varphi:\mathcal{M}_1\to \mathbb{R}^\infty\}$, on a smooth manifold $\mathcal{M}_1$ or
    \item[-] point-like particles, $S_\text{new}^\text{kin}=\{X:\mathbb{R}\to\mathcal{M}_2\}$, on a smooth manifold $\mathcal{M}_2$, or
    \item[-] line-shaped particles, $S_\text{new}^\text{kin}=\{X:\mathbb{R}^2\to\mathcal{M}_3\}$, on a smooth manifold $\mathcal{M}_3$, or
    \item[-] tangent vector fields, $S_\text{new}^\text{kin}=\{\vec{\varphi}:\mathcal{M}_4\to\text{T}\mathcal{M}_4\}$, on a smooth manifold $\mathcal{M}_4$, or
    \item[-] sections, $S_\text{new}^\text{kin}=\{\vec{\varphi}:B_5\to E_5\}$, of a fiber bundle $\langle E_5,B_5,F_5,\pi:E_5\to B_5\rangle$,
\end{itemize}
Spacetime representation theory will help us canvas all of these possibilities.

My approach should be briefly compared with \cite{DeHaro2021} who also employ an analogy with group representation theory in relation to what they call ``dualities''. In particular, they think of dualities as a ``giant symmetry'' which operates one level higher than we are used to; Namely, whereas symmetries are a certain kind of dynamics-preserving maps between a theory's models, dualities are a certain kind of dynamics-preserving maps \textit{between two theories}. But how do their dualities compare to my topological redescriptions? The chief point of contrast is that whereas their dualities are meant to be symmetry-like, my topological redescriptions are meant to be more like coordinate redescriptions (or nomological redescriptions, or geometric redescriptions). More concretely, their symmetry-like dualities are taken to preserve certain aspects of the theory's interpretation \textit{as well as the form} of the theory's dynamical equations. For this reason, they have told me (privately) that they do not consider my above-discussed QHO example to be a duality in their sense.

What tools do we have to help us explore this wide landscape of possible topological framings for our theories? Ideally we would be able to remove and replace the topological underpinnings of our spacetime theories just as easily as we can switch between different coordinate systems (or different law-like axiomatizations, or different geometric framings). As I will now discuss, we can do exactly this with the recently developed ISE Method of topological redescription \citep{GrimmerThesis}. The remainder of this section will be spent introducing a theorem which sets the scope of the ISE Method (for proof see \citeauthor{GrimmerThesis} \citeyear[Ch.11]{GrimmerThesis}). For more information on how the ISE Method works, see \cite{GrimmerThesis}. 

One important caveat in what follows is that (at least for the moment) the ISE Method can only be applied to spacetime theories which are set on smooth manifolds, $\Sigma=\mathcal{M}$. Some more technical work is needed before these techniques can be applied outside of this context, e.g., to theories about the sections, $S_\text{new}^\text{all}=\{\vec{\varphi}:B\to E\}$, of a generic fiber bundle $\Sigma=\langle E,B,F,\pi:E\to B\rangle$.

\begin{figure}[t]
\centering 
\includegraphics[width=0.45\textwidth]{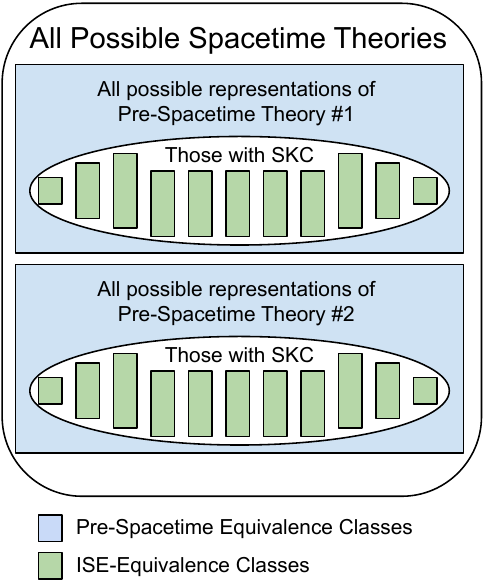} \ 
\includegraphics[width=0.45\textwidth]{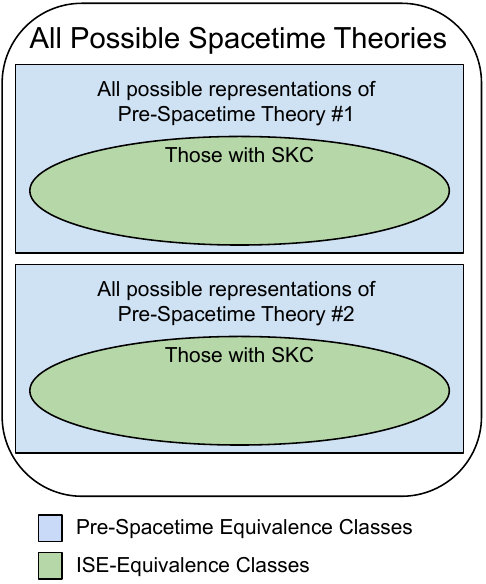}
\caption{Each side of this figure shows all possible spacetime theories being divided up into pre-spacetime equivalence classes (the large blue boxes). Two spacetime theories are in the same blue box if and only if their kinematic and dynamical structures are isomorphic, $\mathcal{S}_\text{new}^\text{kin}\cong \mathcal{S}_\text{old}^\text{kin}$ with $S_\text{new}^\text{dyn}\leftrightarrow S_\text{old}^\text{dyn}$. We can say that any two such theories are different spacetime representations of the same pre-spacetime theoretic structure. The ISE Method applies to the subset of spacetime theories with spacetime-kinematic compatibility (SKC). I have defined this constraint in \citeauthor{GrimmerThesis} (\citeyear[Sec.9.2]{GrimmerThesis}) and argued that it is a weak constraint in \citeauthor{GrimmerThesis} (\citeyear[Sec.9.3]{GrimmerThesis}). In \citeauthor{GrimmerThesis} (\citeyear[Ch.11]{GrimmerThesis}), I have proved that ISE-interrelatedness divides the theories with SKC up into ISE-equivalence classes (the small green boxes, left). Two spacetime theories are in the same green box if and only if they can be redescribed as one another using the ISE Method of topological redescription. Moreover, in \citeauthor{GrimmerThesis} (\citeyear[Ch.11]{GrimmerThesis}) I also proved that these ISE-equivalence classes are as large as possible (the green oval, right). That is, the ISE Method offers us every possible topological description of a given theory, bounded only by a weak spacetime-kinematic compatibility constraint. \textit{(Reproduced from \cite{GrimmerThesis} with permission.)}}\label{FigISEEquivSketch}
\end{figure}

Our above discussion has already divided up the set of all spacetime theories into pre-spacetime equivalence classes, see the large blue boxes in Fig.~\ref{FigISEEquivSketch}. Two spacetime theories are in the same blue box if and only if their kinematic and dynamical structures are isomorphic, $\mathcal{S}_\text{new}^\text{kin}\cong \mathcal{S}_\text{old}^\text{kin}$ with $S_\text{new}^\text{dyn}\leftrightarrow S_\text{old}^\text{dyn}$. The ISE Method applies to the subset of spacetime theories with spacetime-kinematic compatibility (SKC). For the technical definition of this constraint, see \citeauthor{GrimmerThesis} (\citeyear[Sec.9.2]{GrimmerThesis}). For now, just know that it is a very weak constraint as I have argued in \citeauthor{GrimmerThesis} (\citeyear[Sec.9.3]{GrimmerThesis}). In \citeauthor{GrimmerThesis} (\citeyear[Ch.11]{GrimmerThesis}), I have proved that ISE-relatedness is an equivalence relation over the set of all spacetime theories with spacetime-kinematic compatibility. Hence, it divides these theories up into ISE-equivalence classes. See the small green boxes on the left half of  Fig.~\ref{FigISEEquivSketch}. Two spacetime theories are in the same green box if and only if they can be redescribed as one another using the ISE Method of topological redescription. We want these ISE-equivalence classes to be as large as possible such that our one topological redescription technique (the ISE Method) can cover a very wide range of topological redescriptions.

In \citeauthor{GrimmerThesis} (\citeyear[Ch.11]{GrimmerThesis}) I have proved that these ISE-equivalence classes are as large as they can possibly be. Indeed, the small green boxes on the left half of  Fig.~\ref{FigISEEquivSketch} are misleading. Instead, a theory's ISE-equivalence class includes all possible topological redescriptions of that theory which have spacetime-kinematic compatibility. See the green ovals in the right half of Fig.~\ref{FigISEEquivSketch}. Thus, the ISE Method applies to a wide range of spacetime theories and gives us access to effectively every possible topological redescription thereof; It is limited only by a weak spacetime-kinematic compatibility constraint.

\section{Conclusion and Outlook}\label{SecConclusion}
\quad \ \, This paper has introduced spacetime representation theory as a general framework for understanding our capacity to topologically redescribe our spacetime theories. My goal in introducing spacetime representation theory (and the ISE Method) is to gain some mathematical control over the range of all possible spacetime representations of our theories. This should then help us answer some of the interesting philosophical questions which are raised by the possibility of alternative spacetime framings for our theories.

But what philosophical benefits can we expect to follow from our newfound capacity for topological redescription? As I have argued elsewhere (namely, \citeauthor{GrimmerThesis} \citeyear[Ch.3]{GrimmerThesis}), it is reasonable to expect similar benefits to those which have already come about from our ability to redescribe our theories in terms of different axiom-like laws, coordinate systems, and/or geometric structures. Indeed, as I will now overview, these existing capacities for formal redescription have already yielded a wide range of philosophical benefits regarding: theory development, epistemology, metaphysics, and theoretical equivalence/interpretation. It stands to reason that--- given a similar capacity for topological redescription---similar benefits might be achieved in the context of spacetime topology.

Regarding theory development, first note the role that mastering coordinate redescription (and with it tensor calculus) had on the development of general relativity.\footnote{See \cite{Norton1993} for a historical overview.} Moreover, consider how useful our capacity for geometric redescription (e.g., Weyl transformations) was to \cite{EinsteinFokker} as they developed their reformulation of \citeauthor{Nordstrom2}'s (\citeyear{Nordstrom2}) second theory of gravity. This was an important stepping stone on the way to general relativity.\footnote{See \cite{NortonNordstrom}.} In general, we can expect that mastering various kinds of formal redescription will help us to search for new and perhaps better ways of reframing our theories. I claim that developing a strong capacity for topological redescription may be similarly helpful to physicists with their ongoing efforts to develop theories of quantum gravity. I am firmly of the belief, however, that what is needed on this front is not only new mathematical tools, but also improvements in our understanding of the epistemology and metaphysics of space and time. To see how our newfound capacity for topological redescription could help us here, allow me to first sketch how our existing capacities for law-like and geometric redescription have influenced our understanding of their epistemology and metaphysics.

Mastering various kinds of formal redescription might---independent of its potential to help with theory development---also help to shed light on the epistemic process of theory development. For instance, notice how \citeauthor{LewisDavid1983Nwfa}' (\citeyear[pg. 41]{LewisDavid1983Nwfa}) Best Systems Account of laws is built upon an epistemological account of laws: ``I take a suitable system to be one that has the virtues we aspire to in our own theory-building [\dots] it must be as simple in axiomatisation as it can be'' in balance with other desiderata. Note that the feasibility of Lewis' epistemic account of laws rests upon our ability to redescribe our theories using different axiom-like laws. Once we have all of the options on the table, so to speak, we can then ask: Why have our theories come to feature these particular axiom-like laws (or coordinates, geometry, topology, etc.) rather than some mathematically equivalent alternatives? The debates for and against various selection criteria (e.g., best balancing simplicity and strength) are aided by us having a good level of mathematical control over this space of alternative descriptions.

Given that we have a similarly strong capacity for geometric redescription, one can imagine systematically exploring all possible ways of geometrically reframing our spacetime theories. Perhaps surprisingly, there are in this regard many options available to us. Consider the fact that Newton's theory of gravity can be geometrically reformulated as a curved spacetime theory: i.e., Newton-Cartan theory.\footnote{For more information on Newton-Cartan theory, see \citeauthor{MalamentTopics} (\citeyear[Ch.4]{MalamentTopics}). See also the efforts originating from~\cite{SaundersSimon2013RNP} to reframe Newtonian gravity on a Maxwell spacetime.} Moreover, as I noted above, \cite{EinsteinFokker} were able to fruitfully reformulate \citeauthor{Nordstrom2}'s (\citeyear{Nordstrom2}) second theory of gravity using a Weyl transformation. More recently, there has been much discussion of the Geometrical Trinity of Gravity and its non-relativistic counterpart, see \cite{GeometricalTrinity} and \cite{NonRelTrinity} respectively. Given all of these options, the question naturally arises: What selection criteria have led us to favor some of these geometric framings at the expense of some others? Here too, the debates for and against various selection criteria (e.g., \citeauthor{KNOX2019118}'s (\citeyear{KNOX2019118}) spacetime functionalism, or the conventionalism of \cite{pittphilsci22172}) are aided by us having a well-developed capacity for geometric redescription. 

Arguably, developing a similarly strong capacity for topological redescription will similarly aide debates about its epistemology, e.g., its selection criteria. Why have our theories come to feature the particular spacetime topology that they do rather than some pre-spacetime theoretically equivalent alternatives? That is, what are our topology selection criteria? And are these criteria objective? Or is there some element of conventionality to them? In a M\"{o}bius-Euclid world, could it be that one rational civilization opts for the M\"{o}bius representation whereas another opts for the Euclid representation? In our world, could some other civilization experience the world in terms of (what we call) Fourier space?

Returning to the laws' context, Lewis' Best Systems Account of laws is, of course, intended to be a metaphysical account of laws (albeit one which is built upon an epistemological account). In particular, \citeauthor{Hall2015} (\citeyear[pg. 265]{Hall2015}) has identified the  ``unofficial guiding idea'' behind Humeanism about laws as consisting in them ``taking standards that both sides [of the debate] endorse---but that [their anti-Humean] opponent views solely as epistemic standards---and elevating them to the status of standards constitutive of the laws of nature.'' That is, for Humeans the nomological significance of our theories' laws is supposed to be fully accounted for by them winning this best systematization contest. 

We can transfer Hall's unofficial guiding idea from the context of laws to geometry. Namely, given some epistemic account of our geometry selection criteria, we can try to elevate it to a metaphysical account of geometry. This brings us into the orbit of the dynamical vs geometrical spacetime debate. For instance, \cite{HuggettNick2006TRAo} gives a Lewis-inspired implementation of \citeauthor{RBrown2005}'s (\citeyear{RBrown2005}) dynamical approach to spacetime geometry. In \cite{GrimmerThesis}, I have reframed this debate as a competition between the dynamics-first and dynamics-second views of geometry.\footnote{Dynamics-second views of geometry are on the geometrical side of the geometrical vs dynamical spacetime debate. This largely orthodox view is implicit in \citeauthor{Friedman1983}'s textbook (\citeyear[Ch. 6 Sec. 4]{Friedman1983}). It can be seen more explicitly in \cite{MaudlinTim2012Pop:,JANSSEN200926,Norton2008,DoratoMauro2007RTbS}. The dynamics-first views of geometry are, roughly, the dynamical side of the geometrical vs dynamical spacetime debate. For a recent overview of the dynamics-first side of this debate see \cite{BrownRead2018} and references therein, especially \cite{RBrown2005,Nonentity,BrownPooley1999,Menon2019,HuggettNick2006TRAo,StevensSyman2014Tdat}.} Just as in the law's context, the key question here is: Where have the geometric structures which appear in our spacetime theories (e.g., $\eta_\text{ab}$ in special relativity) acquired their (chrono-)geometric significance? Dynamics-second views would claim that they do so by latching onto some dynamics-facilitating geometric counterpart out there in the world. By contrast, dynamics-first views would claim that they gain their geometric significance only by being a particularly nice way of systematizing the dynamical behavior of matter.

Notably, \citeauthor{Norton2008} (\citeyear[pg. 821, 824]{Norton2008}) has complained that dynamical approaches to geometry ``tacitly assume an already existing spacetime endowed with topological properties'' and so ``antecedently presume the essential commitments of a realist conception of spacetime''. Ultimately, \citeauthor{Norton2008}'s (\citeyear[pg. 833]{Norton2008}) problem with this kind of view is that it becomes ``less interesting the more spatiotemporal properties it assumes.'' Addressing this complaint was, in fact, one of my primary motivations in developing the ISE Method of topological redescription in \cite{GrimmerThesis}.

Indeed, once we have an epistemic account of our topology selection criteria, we can then try to elevate this to a metaphysical account (paralleling \citeauthor{Hall2015}'s (\citeyear{Hall2015}) unofficial guiding idea behind Humeanism about laws). Recall from Sec.~\ref{SecIntroduction} that in the law's context the ensuing debate is between dynamics-first and dynamics-second views of laws. The key question is: Where do our theories' laws get their nomological significance from? Is it by helping us systematize dynamics, or by facilitating that dynamics? An analogous debate can now be launched in the context of spacetime topology. 

The prospect of extending certain aspects of Hume's work on laws and causation to the metaphysics of space and time ought to ring some Kantian bells. Indeed, as I have discussed at length in \cite{GrimmerThesis}, the resulting dynamics-first view of spacetime topology has a somewhat Kantian character: The spacetime manifold which appears in our best physical theories does not gain its spatiotemporal significance by latching onto some spatiotemporal stage upon which the world's drama plays out (contra Newton). Nor does it do so by latching onto any fundamental spatiotemporal relationships which help to facilitate interactions between the drama's players (contra Leibniz). Instead, it acquires its spatiotemporal significance only because it is a particularly nice way of systematizing the dynamical behavior of matter.

The final kind of benefit which mastering various kinds of formal redescription can bring relates to issues of theoretical equivalence and interpretation. In fact, while I have carried out the above discussion in terms of our capacity for ``redescription'', it could have equally well been recast in terms of theoretical equivalence. Indeed, mastering a certain kind of formal redescription (e.g., of our theory's laws, coordinates, geometry, topology, etc.) will always come along with a corresponding notion of theoretical equivalence. Namely, we might then say that ``These two models/theories are equivalent in the sense that they are describing the same thing just using different laws, coordinates, geometry, topology, etc.'' These assessments of theoretical equivalence might then guide us in interpreting our theories (e.g., following \citeauthor{DewarInternal}'s (\citeyear{DewarInternal}) discussion of internal interpretations). Indeed, whenever two theories are deemed to be redescriptions of each other the question naturally arises: What exactly is the common core which these two theories are equivalent descriptions of? This question has proved particularly thorny in the context of spacetime dualities.\footnote{See the references listed in footnote \ref{FnDualityWorks}.} The ISE Method is likely to be particularly useful in answering this questions since it allows us to move between the different spacetime framings of our theories through some spacetime-neutral intermediary, namely a pre-spacetime theory.



\singlespacing
\bibliographystyle{dcu}
\bibliography{Bibliography}

@article{Norton1993,
	doi = {10.1088/0034-4885/56/7/001},
	year = {1993},
	month = {jul},
	publisher = {{IOP} Publishing},
	volume = {56},
	number = {7},
	pages = {791--858},
	author = {J D Norton},
        title = {{General Covariance and the Foundations of General Relativity: Eight Decades of Dispute}},	
        journal = {Reports on Progress in Physics}
}

@book{Friedman1983,
 author = {Michael Friedman},
 publisher = {Princeton University Press},
 address = {Princeton, NJ},
 title = {Foundations of Space-Time Theories: Relativistic Physics and Philosophy of Science},
 urldate = {2024-05-15},
 year = {1983}
}

@article{Norton2008,
issn = {0007-0882},
journal = {The British Journal for the Philosophy of Science},
pages = {821--834},
volume = {59},
publisher = {Oxford University Press},
address = {Oxford, UK},
number = {4},
year = {2008},
title = {{Why Constructive Relativity Fails}},
author = {Norton, John D.},
}

@article{Kempf_2010,
	doi = {10.1088/1367-2630/12/11/115001},
	year = 2010,
	month = {nov},
	publisher = {{IOP} Publishing},
	volume = {12},
	number = {11},
	pages = {e115001},
	author = {Achim Kempf},
	title = {{Spacetime Could Be Simultaneously Continuous And Discrete, In The Same Way That Information Can Be}},
	journal = {New Journal of Physics}
}

@Article{Kempf2018,
author={Kempf, Achim},
title={{Quantum Gravity, Information Theory and the CMB}},
journal={Foundations of Physics},
year={2018},
month={Oct},
day={01},
volume={48},
number={10},
pages={1191-1203},
issn={1572-9516},
doi={10.1007/s10701-018-0163-2},
}

@article{Menon2019,
issn = {0031-8248},
journal = {Philosophy of Science},
pages = {1273--1283},
volume = {86},
publisher = {The University of Chicago Press},
number = {5},
year = {2019},
title = {{Algebraic Fields and the Dynamical Approach to Physical Geometry}},
copyright = {Copyright 2019 by the Philosophy of Science Association. All rights reserved.},
language = {eng},
address = {Chicago, IL},
author = {Menon, Tushar},
}

@article{Nonentity,
  edition = {},
  number = {C},
  journal = {Philosophy and Foundations of Physics},
  pages = {67-89},
  publisher = {},
  school = {},
  title = {{Minkowski Space-Time: A Glorious Non-Entity}},
  volume = {1},
  author = {Brown, H and Pooley, O},
  editor = {},
  year = {2006},
  series = {}
}

@incollection{BrownPooley1999,
 author = {Pooley, Oliver and Brown, Harvey},
 booktitle = {Physics Meets Philosophy at the Planck Scale},
 editor = {Callender, Craig and Huggett, Nick},
 note = {256--272},
 publisher = {Cambridge University Press},
 title = {{The Origins of the Spacetime Metric: Bell’s Lorentzian Pedagogy and its Significance in General Relativity}},
 year = {1999},
 address = {Cambridge, UK}
}

@phdthesis{StevensSyman2014Tdat,
publisher = {ProQuest Dissertations Publishing},
year = {2014},
title = {{The Dynamical Approach to Relativity as a Form of Regularity Relationalism}},
author = {Stevens, Syman},
address = {Oxford, UK},
school = {University of Oxford},
}

@article{HuggettNick2006TRAo,
issn = {0026-4423},
journal = {Mind},
pages = {41--73},
volume = {115},
publisher = {Oxford University Press},
address = {Oxford, UK},
number = {457},
year = {2006},
title = {{The Regularity Account of Relational Spacetime}},
author = {Huggett, Nick},
}

@article{DoratoMauro2007RTbS,
issn = {0269-8595},
journal = {{International Studies in the Philosophy of Science}},
pages = {95--102},
volume = {21},
publisher = {Routledge},
number = {1},
year = {2007},
title = {{Relativity Theory between Structural and Dynamical Explanations}},
language = {eng},
author = {Dorato, Mauro},
}

@article{LewisDavid1983Nwfa,
issn = {0004-8402},
journal = {Australasian Journal of Philosophy},
pages = {343--377},
volume = {61},
publisher = {Taylor & Francis Group},
number = {4},
year = {1983},
title = {{New Work for a Theory of Universals}},
language = {eng},
author = {Lewis, David},
}

@book{RBrown2005,
	year = {2005},
	publisher = {Oxford University Press},
        address = {Oxford, UK},
	title = {Physical Relativity: Space-Time Structure From a Dynamical Perspective},
	author = {Harvey R. Brown}
}

@book{MaudlinTim2012Pop:,
series = {Princeton Foundations of Contemporary Philosophy},
publisher = {Princeton University Press},
booktitle = {{Philosophy of Physics: Space and Time}},
isbn = {9781400842339},
year = {2012},
title = {{Philosophy of Physics: Space and Time}},
address = {Princeton, NJ},
author = {Maudlin, Tim},
}

@incollection{BrownRead2018,
language = {eng},
note = {70-85},
pagination = {none},
publisher = {Routledge},
address = {London, UK},
title = {{The Dynamical Approach to Spacetime Theories}},
year = {2022},
editor = {Knox, Eleanor and Wilson, Alastair},
author = {Brown, Harvey R. and Read, James},
booktitle = {The Routledge Companion to Philosophy of Physics},
copyright = {2022 Taylor & Francis},
edition = {1st},
isbn = {1138653071},
}

@incollection{Hall2015,
author = {Hall, Ned},
publisher = {John Wiley and Sons, Ltd},
isbn = {9781118398593},
title = {{Humean Reductionism about Laws of Nature}},
booktitle = {A Companion to David Lewis},
editor = {Loewer, Barry and Schaffer, Jonathan},
address = {Chichester, UK},
note = {262-277},
doi = {https://doi.org/10.1002/9781118398593.ch17},
eprint = {https://onlinelibrary.wiley.com/doi/pdf/10.1002/9781118398593.ch17},
year = {2015}
}

@article{KNOX2019118,
title = {{Physical Relativity from a Functionalist Perspective}},
journal = {Studies in History and Philosophy of Science Part B: Studies in History and Philosophy of Modern Physics},
volume = {67},
pages = {118-124},
year = {2019},
issn = {1355-2198},
doi = {https://doi.org/10.1016/j.shpsb.2017.09.008},
author = {Eleanor Knox}
}

@article{JANSSEN200926,
title = {{Drawing the Line between Kinematics and Dynamics in Special Relativity}},
journal = {Studies in History and Philosophy of Science Part B: Studies in History and Philosophy of Modern Physics},
volume = {40},
number = {1},
pages = {26-52},
year = {2009},
issn = {1355-2198},
doi = {https://doi.org/10.1016/j.shpsb.2008.06.004},
author = {Michel Janssen}
}

@book{MalamentTopics,
series = {Chicago Lectures in Physics},
publisher = {University of Chicago Press},
address = {Chicago, IL},
booktitle = {{Topics in the Foundations of General Relativity and Newtonian Gravitation Theory}},
year = {2012},
title = {{Topics in the Foundations of General Relativity and Newtonian Gravitation Theory}},
language = {eng},
author = {Malament, David B},
}

@Article{Beem1991,
author={Beem, John K.
and Parker, Phillip E.},
title={{The Space of Geodesics}},
journal={Geometriae Dedicata},
year={1991},
month={Apr},
day={01},
volume={38},
number={1},
pages={87-99},
issn={1572-9168},
doi={10.1007/BF00147737},
}

@article{BH2016,
issn = {1755-0203},
pages = {556--582},
volume = {9},
publisher = {Cambridge University Press},
journal = {The Review of Symbolic Logic},
number = {3},
year = {2016},
title = {{Morita Equivalence}},
author = {Barrett, Thomas William and Halvorson, Hans},
}

@article{Nordstrom2,
    author = {Nordstr\"{o}m, Gunnar},
    title = {{Zur Theorie der Gravitation vom Standpunkt der Relativitätsprinzips}},
    journal = {Annalen der Physik},
    year = {1913},
    volume = {42},
    pages = {533-554},
}

@article{EinsteinFokker,
issn = {0003-3804},
journal = {Annalen der Physik},
language = {ger},
number = {10},
author = {Einstein, A. and Fokker, A. D.},
address = {Leipzig},
title = {{Die Nordströmsche Gravitationstheorie vom Standpunkt des absoluten Differentialkalküls}},
volume = {349},
year = {1914},
pages = {321-328},
publisher = {WILEY-VCH Verlag},
}

@Article{DeHaro2021,
author={De Haro, Sebastian
and Butterfield, Jeremy},
title={{On Symmetry and Duality}},
journal={Synthese},
year={2021},
month={Apr},
day={01},
volume={198},
number={4},
pages={2973-3013},
issn={1573-0964},
doi={10.1007/s11229-019-02258-x},
}

@book{NaturalGaugeNatural,
  title={Natural and Gauge Natural Formalism for Classical Field Theorie: A Geometric Perspective including Spinors and Gauge Theories},
  author={Fatibene, L. and Francaviglia, M.},
  isbn={9781402017032},
  lccn={03064183},
  year={2003},
  publisher={Springer},
  address={Dordrecht, NL}
}

@article{SaundersSimon2013RNP,
language = {eng},
number = {1},
pages = {22-48},
publisher = {University of Chicago Press},
title = {{Rethinking Newton's Principia}},
volume = {80},
year = {2013},
author = {Saunders, Simon},
address = {Chicago, IL},
copyright = {Copyright 2013 by the Philosophy of Science Association. All rights reserved.},
issn = {0031-8248},
journal = {Philosophy of Science},
}

@phdthesis{GrimmerThesis,
author={Grimmer, Daniel},
title={Searching for New Spacetimes: Towards a Dynamics-First View of Topology},
year={2024},
school={University of Oxford},
address = {Oxford, UK}
}

@article{NortonNordstrom,
author = {Norton, John D.},
address = {NEW YORK},
copyright = {1993 INIST-CNRS},
issn = {0003-9519},
journal = {Archive for history of exact sciences},
language = {eng},
number = {1},
pages = {17-94},
title = {{Einstein, Nordström and the Early Demise of Scalar, Lorentz-Covariant Theories of Gravitation}},
volume = {45},
year = {1992},
publisher = {Springer-Verlag},
}

@misc{pittphilsci22172,
            year = {2023},
          author = {Patrick D{\"u}rr and James Read},
           title = {{Reconsidering Conventionalism: An Invitation to a Sophisticated Philosophy for Modern (Space-)Times}},
           month = {May},
        note = {Phil. Sci. Archive 22172},
}

@Article{GeometricalTrinity,
AUTHOR = {Beltrán Jiménez, Jose and Heisenberg, Lavinia and Koivisto, Tomi S.},
TITLE = {{The Geometrical Trinity of Gravity}},
JOURNAL = {Universe},
VOLUME = {5},
YEAR = {2019},
NUMBER = {7},
ARTICLE-NUMBER = {173},
ISSN = {2218-1997},
DOI = {10.3390/universe5070173}
}

@misc{NonRelTrinity,
      title={{The Non-Relativistic Geometric Trinity of Gravity}}, 
      author={William J. Wolf and James Read and Quentin Vigneron},
      year={2024},
      eprint={2308.07100},
      archivePrefix={arXiv},
      primaryClass={gr-qc},
      note={Pre-print ArXiv:2308.07100},
}

@article{DewarInternal,
	author = {Neil Dewar},
	doi = {10.1007/s11229-023-04102-9},
	journal = {Synthese},
	number = {4},
	pages = {1--24},
	publisher = {Springer Verlag},
	title = {{Interpretation and Equivalence; or, Equivalence and Interpretation}},
	volume = {201},
	year = {2023}
}

@article{RasmusEnrico,
author = {Grimmer, Daniel and Cinti, Enrico and Jaksland, Rasmus},
title = {{Duality, Underdetermination, and the Uncommon Common Core}},
journal = {The British Journal for the Philosophy of Science},
volume = {0},
year = {2024},
doi = {10.1086/730421},
eprint = { https://doi.org/10.1086/730421}
}

@article{de_haro_theoretical_2019,
	title = {{Theoretical Equivalence and Duality}},
	issn = {1573-0964},
	doi = {10.1007/s11229-019-02394-4},
	journal = {Synthese},
	author = {De Haro, Sebastian},
	month = sep,
	year = {2019},
}

@article{de_haro_empirical_2023,
	title = {{The Empirical Under-Determination Argument Against Scientific Realism for Dual Theories}},
	volume = {88},
	issn = {1572-8420},
	doi = {10.1007/s10670-020-00342-0},
	pages = {117--145},
	number = {1},
	journal = {Erkenntnis},
	shortjournal = {Erkenntnis},
	author = {De Haro, Sebastian},
	year = {2023},
}

@incollection{de_haro_schema_2018,
	location = {Cham},
	title = {{A Schema for Duality, Illustrated by Bosonization}},
	isbn = {978-3-319-64813-2},
	pages = {305--376},
	booktitle = {Foundations of Mathematics and Physics One Century After Hilbert: New Perspectives},
	publisher = {Springer International Publishing},
	author = {De Haro, Sebastian and Butterfield, Jeremy},
	editor = {Kouneiher, Joseph},
	year = {2018},
}

@article{de_haro_symmetry_2021,
	title = {{On Symmetry and Duality}},
	volume = {198},
	issn = {1573-0964},
	doi = {10.1007/s11229-019-02258-x},
	pages = {2973--3013},
	number = {4},
	journal = {Synthese},
	shortjournal = {Synthese},
	author = {De Haro, Sebastian and Butterfield, Jeremy},
	year = {2021},
}

@book{deharo2023,
  title={{The Philosophy and Physics of Duality}},
  author={De Haro, Sebastian and Butterfield, Jeremy},
  year={Forthcoming},
  publisher={Oxford University Press}
}

@article{de_haro_comparing_2017,
	title = {{Comparing Dualities and Gauge Symmetries}},
	volume = {59},
	pages = {68--80},
	journal = {Studies in History and Philosophy of Science Part B: Studies in History and Philosophy of Modern Physics},
	shortjournal = {Studies in History and Philosophy of Science Part B: Studies in History and Philosophy of Modern Physics},
	author = {De Haro, Sebastian and Teh, Nicholas and Butterfield, Jeremy N.},
	year = {2017},
}

@article{huggett_target_2017,
	title = {{Target Space is Not Equal to Space}},
	volume = {59},
	pages = {81--88},
	journal = {Studies in History and Philosophy of Science Part B: Studies in History and Philosophy of Modern Physics},
	shortjournal = {Studies in History and Philosophy of Science Part B: Studies in History and Philosophy of Modern Physics},
	author = {Huggett, Nick},
	year = {2017},
}

@article{huggett_emergent_2013,
	title = {{Emergent Spacetime and Empirical (In)coherence}},
	volume = {44},
	number = {3},
	journal = {Studies in History and Philosophy of Science Part B: Studies in History and Philosophy of Modern Physics},
	author = {Huggett, Nick and Wüthrich, Christian},
	year = {2013},
	pages = {276--285},
}

@article{jaksland_holography_2020,
	title = {{Holography without Holography: How to Turn Inter-representational into Intra-theoretical Relations in AdS/CFT}},
	volume = {71},
	issn = {1355-2198},
	doi = {10.1016/j.shpsb.2020.04.007},
	journal = {Studies in History and Philosophy of Science Part B: Studies in History and Philosophy of Modern Physics},
	author = {Jaksland, Rasmus and Linnemann, Niels S.},
	month = aug,
	year = {2020},
	pages = {101--117},
}

@article{le_bihan_duality_2018,
	title = {{Duality and Ontology}},
	volume = {13},
	pages = {e12555},
	number = {12},
	journal = {Philosophy Compass},
	shortjournal = {Philosophy Compass},
	author = {Le Bihan, Baptiste and Read, James},
	urldate = {2019},
	year = {2018},
}

@article{matsubara_realism_2013,
	title = {{Realism, Underdetermination and String Theory Dualities}},
	volume = {190},
	number = {3},
	journal = {Synthese},
	author = {Matsubara, Keizo},
	year = {2013},
	pages = {471--489},
}

@article{read_motivating_2018,
	title = {{Motivating Dualities}},
	issn = {1573-0964},
	doi = {10.1007/s11229-018-1817-5},
	journal = {Synthese},
	shortjournal = {Synthese},
	author = {Read, James and Møller-Nielsen, Thomas},
	year = {2018},
}

@article{rickles_dual_2017,
	title = {{Dual theories: ‘Same but Different’ or ‘Different but Same’?}},
	pages = {62--67},
	journal = {Studies in History and Philosophy of Science Part B: Studies in History and Philosophy of Modern Physics},
	shortjournal = {Studies in History and Philosophy of Science Part B: Studies in History and Philosophy of Modern Physics},
	author = {Rickles, Dean},
	year = {2017},
}

\end{document}